\documentclass[amsmath,superscriptaddress, amssymb,aps,prd,reprint,longbibliography,nofootinbib,notitlepage,preprintnumbers,10pt]{revtex4-2}
\usepackage{natbib}

\RequirePackage[colorlinks,citecolor=blue,urlcolor=magenta,linkcolor=blue]{hyperref}
\usepackage{graphicx}
\usepackage{enumerate}
\usepackage[shortlabels]{enumitem}
\usepackage{url}
\usepackage{xcolor}
\usepackage{graphicx}
\usepackage{setspace}
\usepackage{dutchcal}
\usepackage{tabularx}
\usepackage{scalerel}
\usepackage{braket}
\usepackage{tikz}
\usepackage{tikz-cd}
\usetikzlibrary{svg.path}

\newcommand{\rd}{{\rm d}}

\newcommand{\e}{{\rm e}}

\usepackage{moreenum}
\usepackage[utf8]{inputenc}

\labelformat{equation}{Eq.~(#1)} 
\labelformat{figure}{Fig.~#1} 
\labelformat{subfigure}{Fig.~\thefigure#1} 
\labelformat{table}{Tab.~#1} 
\labelformat{appendix}{Appendix #1}

\definecolor{orcidlogocol}{HTML}{A6CE39}
\tikzset{
  orcidlogo/.pic={
    \fill[orcidlogocol] svg{M256,128c0,70.7-57.3,128-128,128C57.3,256,0,198.7,0,128C0,57.3,57.3,0,128,0C198.7,0,256,57.3,256,128z};
    \fill[white] svg{M86.3,186.2H70.9V79.1h15.4v48.4V186.2z}
                 svg{M108.9,79.1h41.6c39.6,0,57,28.3,57,53.6c0,27.5-21.5,53.6-56.8,53.6h-41.8V79.1z M124.3,172.4h24.5c34.9,0,42.9-26.5,42.9-39.7c0-21.5-13.7-39.7-43.7-39.7h-23.7V172.4z}
                 svg{M88.7,56.8c0,5.5-4.5,10.1-10.1,10.1c-5.6,0-10.1-4.6-10.1-10.1c0-5.6,4.5-10.1,10.1-10.1C84.2,46.7,88.7,51.3,88.7,56.8z};
  }
}

\newcommand\orcidicon[1]{\href{https://orcid.org/#1}{\mbox{\scalerel*{
\begin{tikzpicture}[yscale=-1,transform shape]
\pic{orcidlogo};
\end{tikzpicture}
}{|}}}}

\usepackage{hyperref}

\usepackage{bm}
\usepackage{xcolor}
\usepackage{bm}
\usepackage{slashed}
\usepackage{xparse}
\usepackage{dsfont}
\usepackage[mathscr]{euscript}
\usepackage{float}
\usepackage{etoolbox}
\patchcmd{\section}
  {\centering}
  {\raggedright}
  {}
  {}
\patchcmd{\subsection}
  {\centering}
  {\raggedright}
  {}
  {}
\DeclareUnicodeCharacter{2212}{-}
\allowdisplaybreaks
\begin{document}
\preprint{YITP-25-77, IPMU25-0027}
\title{Anisotropic quantum universe in Ho\v{r}ava-Lifshitz gravity}
\author{Vikramaditya Mondal}
\email{vikramaditya.academics@gmail.com}
\affiliation{School of Physical Sciences, Indian Association for the Cultivation of Science, Kolkata-700032, India}

\author{Shinji Mukohyama}
\email{shinji.mukohyama@yukawa.kyoto-u.ac.jp}
\affiliation{Center for Gravitational Physics and Quantum Information, Yukawa Institute for Theoretical Physics, Kyoto University, 606-8502, Kyoto, Japan}
\affiliation{Kavli Institute for the Physics and Mathematics of the Universe (WPI), The University of Tokyo Institutes for Advanced Study (UTIAS), The University of Tokyo, Kashiwa, Chiba 277-8583, Japan}



\begin{abstract}
We quantize a Bianchi IX universe in Ho\v{r}ava-Lifshitz theory. For analytical tractability, we consider the small anisotropy limit of the Bianchi IX, that is, a perturbative anisotropic deformation of a closed, homogeneous and isotropic universe. In the case of the projectable theory we further set the ``dark matter as integration constant'' to zero by assuming that the space consists of only one connected piece. In that limit and under the assumption, we first study the semi-classical WKB solutions to the Wheeler-DeWitt equation. We find the wave function of the universe, up to an overall normalization, and estimate the semi-classical tunneling probability for the emergence of an expanding universe. We establish a dictionary of correspondence between the WKB wave functions in General Relativity and Ho\v{r}ava-Lifshitz theory in the large-scale factor (or IR) limit. For a small universe (UV limit), on the other hand, due to contributions from higher-dimensional operators, the anisotropies decouple from the scale factor, a behavior significantly different from General Relativity, and analytic solutions to the Wheeler-DeWitt equation beyond the WKB approximation can be found. The wave function of the scale factor satisfies the DeWitt criterion, whereas the wave functions of anisotropies resemble those of quantum harmonic oscillators. The quantum prediction for the initial condition of anisotropies is obtained in terms of the coupling parameters of Ho\v{r}ava-Lifshitz theory. We find a bound on the coupling parameters from the normalizability of the wave functions of anisotropies. Further, we calculate the expectation values for squared anisotropic shear and squared anisotropies in both the large universe and small universe limits.
\end{abstract}

\maketitle


\section{Introduction}
The Cosmic Microwave Background Radiation (CMBR) appears to be incredibly homogeneous and isotropic \cite{Planck:2019evm}, the temperature deviating no more than $10^{-4} \, {\rm K}$ across the sky \cite{Fixsen:2009ug}. As the anisotropies dilute with the expansion of the universe, it is reasonable to think that for the early phases of its evolution, anisotropies did play an important role, even if the present universe seems isotropic \cite{Belinsky:1970ew}. The question is, how much anisotropies were present near the beginning of the universe, that is, what is the initial condition for anisotropies?

The question may be answered by starting from present-day values for anisotropies and running backward in time, and then asking what should have been the initial condition at an initial time, say the Planck time. Even though it is a reasonable approach in cosmology, there remain several concerns. First, since the anisotropies increase towards the past, a tiny error in the measurement of the present anisotropies may lead to enormous error at the early time. Second, at Planckian energy densities, the classical theories cease to be trustworthy since quantum gravitational effects start to become important. Then the explanation for why such an initial condition is generic must be found within the framework of quantum gravity.

Quantum cosmology, which treats the entire universe as a quantum entity, attempts to answer questions regarding the initial conditions. In this approach, the state of the universe is given by a wave function, which provides a probabilistic distribution for the initial conditions of the universe, thereby predicting their most likely values. Therefore, it is important to construct the wave function of an anisotropic universe in order to understand the role of anisotropies in the early universe. In this paper, we will deal with the construction of the wave function of a Bianchi IX universe in Ho\v{r}ava-Lifshitz theory of gravity, and implications thereof.

One of the approaches to quantum cosmology is based on the canonical quantum gravity \cite{DeWitt:1967yk}, in which the Hamiltonian constraint appearing in the Arnowitt, Deser, and Misner (ADM) formulation of General Relativity \cite{Arnowitt:1962hi} is promoted to an operator $\hat{\mathscr{H}}[h]$ and imposed on physical states as a differential equation $\hat{\mathscr{H}}[h] \Psi[h]=0$, where $h_{ij}$ is the induced metric on spatial hypersurfaces of constant time, and $\Psi[h]$ is the wave function(al) living in the configuration space of all 3-geometries (modulo spatial diffeomorphisms) called the \textit{superspace}\footnote{Not to be confused with the ``superspace'' in the theory of Supergravity.}. The dimension of the superspace is infinite, and the differential equation, called the Wheeler-DeWitt equation, is incredibly difficult to solve, see for example \cite{Chakraborty:2023yed}. However, when applied to the cosmological scenarios, the assumption that the early universe is homogeneous reduces the number of degrees of freedom and results in a finite-dimensional configuration space---\textit{minisuperspace}. In such a minisuperspace setup, the Wheeler-DeWitt equation can be solved, if not only approximately, in many cases of interest.

An alternative to the canonical formalism is the path integral formalism \cite{Hartle:1983ai,Feldbrugge:2017kzv,DiTucci:2019dji,di2019noprescription}, which defines the wave function as $\Psi[g]=\int \mathscr{D}{}g^{(4)} \, \e^{\frac{i}{\hbar}S[g^{(4)}]}$, where constraints in the theory is incorporated in the path integral utilizing the BRST formalism \cite{Halliwell:1988wc,Halliwell:1990qr}. The boundary conditions prescribed in the path integral formalism pick out particular solutions of the Wheeler-DeWitt equation. In the present paper, however, we will not explore the path integral formalism.

Since General Relativity is \textit{not} perturbatively renormalizable \cite{tHooft:1974toh,Goroff:1985sz}, one must look for other avenues to determine the initial conditions of the universe. We explore here the consequences for the Ho\v{r}ava-Lifshitz theory of gravity \cite{Horava:2009uw}. There are mainly two versions of Ho\v{r}ava-Lifshitz gravity: projectable and non-projectable ones. For homogeneous universes the only difference between these two versions is that the former admits ``dark matter as integration constant''~\cite{Mukohyama:2009mz} while the latter does not. However, if we assume that the space consists of only one connected piece then for a homogeneous universe the ``dark matter as integration constant'' vanishes and there is no difference between them. In the present paper we make this assumption and work in the projectable version for concreteness. 

The projectable version of Ho\v{r}ava-Lifshitz gravity is perturbatively renormalizable \cite{Barvinsky:2015kil} and has been shown to be asymptotically free \cite{Barvinsky:2023uir}. Previously, it has been shown that in the Ho\v{r}ava-Lifshitz theory, the higher-dimensional operators, which are required for renormalizability, also provides a novel mechanism to produce a scale-invariant power spectrum for the primordial cosmological perturbations \cite{Mukohyama:2009gg} even without inflation. Moreover, in the Ho\v{r}ava-Lifshitz theory, the DeWitt boundary condition---vanishing of the wave function at the classical singular configuration of the universe---has been shown to hold, whereas General Relativity seems to be inconsistent with the DeWitt condition under inhomogeneous perturbations \cite{Matsui:2021yte}. Also, Ho\v{r}ava-Lifshitz theory provides a rich avenue for cosmological phenomenology \cite{Mukohyama:2010xz}. Due to these reasons, we believe it important to consider the Ho\v{r}ava-Lifshitz theory for an investigation into the initial condition for anisotropies.

The investigation of Bianchi IX spacetime in quantum cosmology based on General Relativity has been taken up initially by Misner \cite{Misner:1969ae}. Later, this anisotropic model in quantum cosmology has also been studied in the context of the Hartle-Hawking no-boundary proposal in \cite{Hawking:1984wn} and Vilenkin's tunneling proposal in \cite{delCampo:1989hy}.

In the following, we will first consider the General Relativity case, by reproducing some of the results in \cite{Hawking:1984wn,delCampo:1989hy}, before turning to the Ho\v{r}ava-Lifshitz theory so that the results can be compared. We will write down, for a Bianchi IX universe with positive cosmological constant, the Wheeler-DeWitt equation, which is then solved under the WKB approximation, treating the anisotropies as small perturbations. The superpotential of an isotropic universe has one turning point determined by the cosmological constant, where the WKB solution is not valid. The solutions on both sides of the turning point have to be matched by the WKB connection formula. Specific choices for the modes in these two regions lead to either the Hartle-Hawking no-boundary wave function or Vilenkin's tunneling proposal.

On the other hand, we will consider the particular cases of the Ho\v{r}ava-Lifshitz theory wherein there are three distinct turning points of the (isotropic) superpotential determined by the cosmological constant, and the coupling parameters associated with the higher-dimensional operators. In this case as well, the solutions across the turning points have to be matched. Due to the complicated structure of the isotropic superpotential, an analytical solution for the anisotropic wave function can not be determined except in either small or large universe limits. For intermediate scales, we perform a numerical analysis. We find a correspondence between the wave function in the Ho\v{r}ava-Lifshitz theory and the Hartle-Hawking no-boundary wave function in the large universe limit. Unlike the case of General Relativity, in Ho\v{r}ava-Lifshitz theory we can also determine the wave function of the anisotropic universe in the small universe (or UV) limit, which will provide the initial condition for anisotropies, depending on the coupling parameters of the theory.

The paper is organized as follows: in section \ref{sec:GR_wave_function} we discuss the wave function of a Bianchi IX universe with a positive cosmological constant in General Relativity. In section \ref{sec:HL_wave_function} we start discussing the Ho\v{r}ava-Lifshitz theory. Specifically, in section \ref{sec:HL_wkb_setup} we set up the WKB equations in orders of $\hbar$ and anisotropies to be solved. In sections \ref{sec:wkb_small} and \ref{sec:wkb_hl_large}, we discuss the solutions to these equations in the small and large universe limits, respectively. In section \ref{sec:numerical}, we discuss the numerical solution for the wave functions of anisotropies in the intermediate scale. Finally, in section \ref{sec:initial_conditions}, we discuss the initial conditions for anisotropies and the behavior of anisotropic shear in different regimes.

\section{Wave function of an anisotropic universe in General Relativity}\label{sec:GR_wave_function}
General Relativity is described by the Einstein-Hilbert action
\begin{align}
    S_{\text{E-H}} & = \frac{1}{2\kappa^2}\int \rd^4 x \sqrt{-g} \left(R-2\Lambda\right).
\end{align}
Here $g_{\mu\nu}$ denotes the metric on the spacetime manifold and $g$ is its determinant, $R$ is the corresponding  Ricci scalar, $\Lambda$ is the cosmological constant, and $\kappa^2 = M_{\rm Pl}^{-2}=(8\pi G_{\rm N})$, with $M_{\rm Pl}$, $G_{\rm N}$ being the Planck mass and Newtonian gravitational constant, respectively. In the ADM formulation, the space-time metric is decomposed as
\begin{align}
    \rd s^2 = - N^2 \rd t^2 + h_{ij}(N^{i}\rd t + \rd x^i) (N^{j}\rd t + \rd x^j),
\end{align}
where $N$ is the lapse function and $N^{i}$ is the shift vector, and $h_{ij}$ are induced 3-metrics on the foliated spatial hypersurfaces. Under this decomposition, the above action can be rewritten in the following form
\begin{align}
    S_{\text{E-H}} & = \frac{1}{2\kappa^2} \int \rd t \rd^3 x \sqrt{h} N \left(\mathscr{K}_{ij}\mathscr{K}^{ij} - \mathscr{K}^2 + \mathscr{R} - 2 \Lambda  \right),
\end{align}
up to boundary terms. Here, $h$ is the determinant of the induced 3-metric, $\mathscr{K}_{ij}$ is the extrinsic curvature, and $\mathscr{K}$ its trace, and $\mathscr{R}$ is the 3-Ricci scalar of the spatial hypersurface. The extrinsic curvature is defined as {\sloppy $\mathscr{K}_{ij}=\frac{1}{2N}\left(\partial_{t}h_{ij} - {}^{(3)}\nabla_{i}N_{j} - {}^{(3)}\nabla_{j}N_{i}\right)$}, where ${}^{(3)}\nabla_{i}$ is the covariant derivative on the spatial hypersurface compatible with the induced 3-metric and $N_i=h_{ij}N^j$. In the following, we will evaluate this action for a Bianchi IX anisotropic cosmological spacetime.

The Bianchi IX anisotropic metric is given by
\begin{align}\label{eq:bianchi_ix}
    \rd s^2 = - N^2 \rd t^2 + \frac{a^2}{4} \e^{2\beta_{ij}} \omega^{i} \omega^{j},
\end{align}
where $\omega^i$ and $\beta_{ij}$ are defined as
\begin{subequations}
\begin{align}
    \omega^{1} &= -\sin x^{3} \rd x^{1} + \sin x^{1} \cos x^{3} \rd x^{2},\\
    \omega^{2} &= \cos x^{3} \rd x^{1} + \sin x^{1} \sin x^{3} \rd x^{2},\\
    \omega^{3} &= \cos x^{1} \rd x^{2} + \rd x^{3},\\
    \beta_{ij} & = {\rm diag}(\beta_{+} + \sqrt{3}\beta_{-}, \beta_{+} - \sqrt{3}\beta_{-},-2\beta_{+}).
\end{align} 
\end{subequations}
and the shift vector vanishes $N^{i}=0$.
In this coordinate system, the Einstein-Hilbert action takes the form (up to boundary terms)
\begin{align}\label{eq:einstein-hilbert}
    S_{\text{E-H}} = & \frac{\pi^2}{4\kappa^2} \int \rd t \Bigg[-\frac{3a \dot{a}^2}{N} + \frac{3a^3}{N} (\dot{\beta}_{+}^2 + \dot{\beta}_{-}^2)  \nonumber\\
    & - N \Bigg(\Lambda a^3 + 64 V_{\rm IX}(\beta_{+},\beta_{-}) a \Bigg) \Bigg],
\end{align}
where $V_{\rm IX}$ is the anisotropy potential
\begin{align}
    V_{\rm IX}(\beta_{+},\beta_{-}) = & -\frac{1}{16} \left(\frac{1}{2}\e^{4\beta_{+}}\left(1-\cosh 4\sqrt{3}\beta_{-}\right) \right. \nonumber\\
    & \left. +\e^{-2\beta_{+}}\cosh 2\sqrt{3}\beta_{-} -\frac{1}{4}\e^{-8\beta_{+}}\right).
\end{align}
From the action, we can identify the canonically conjugate momenta as
\begin{align}\label{eq:conjugate_momenta}
    p_{a} = - \frac{3\pi^2}{2\kappa^2} \frac{a\dot{a}}{N},~~~~ p_{\beta_{\pm}} = \frac{3\pi^2}{2\kappa^2}\frac{a^3 \dot{\beta}_{\pm}}{N}.
\end{align}
Using these definitions, the action can be recast into the following form
\begin{align}
    S_{\text{E-H}} = \int \rd t \left[p_{a} \dot{a} + p_{\beta_{+}} \dot{\beta}_{+} + p_{\beta_{-}} \dot{\beta}_{-} - N\mathscr{H}\right],
\end{align}
where $N\mathscr{H}$ is the Hamiltonian of the system defined as
\begin{align}
    N\mathscr{H} = & p_{a} \dot{a} + p_{\beta_{+}} \dot{\beta}_{+} + p_{\beta_{-}} \dot{\beta}_{-} - \mathscr{L} \nonumber\\
    = & N\left[ - \frac{\kappa^2}{3\pi^2} \frac{p_{a}^2}{a} + \frac{\kappa^2}{3\pi^2a^3} \left(p_{\beta_{+}}^2 + p_{\beta_{-}}^2\right)  \right. \nonumber\\
    &  \left. + \frac{\pi^2}{\kappa^2} \Bigg(\frac{\Lambda}{4}a^3 + 16 V_{\rm IX}(\beta_{+},\beta_{-}) a \Bigg)\right].
\end{align}
Varying the action with respect to $N$ leads to the well-known Hamiltonian constraint of General Relativity $\mathscr{H}\approx 0$. We can, then, canonically quantize the system by promoting the conjugate pairs to operators so that and the momenta take the form $p_{a} \to - i \hbar \partial_{a}$ and $p_{\beta_{\pm}} \to - i \hbar \partial_{\beta_{\pm}}$. The Wheeler-DeWitt equation can then be written as follows
\begin{align}
    & \Bigg[ \frac{\kappa^4\hbar^2}{3\pi^4} a^{-p}\frac{\partial}{\partial a}a^{p} \frac{\partial}{\partial a} - \frac{\kappa^4 \hbar^2}{3\pi^4 a^2} \left(\frac{\partial^2}{\partial \beta_{+}^2} + \frac{\partial^2}{\partial \beta_{-}^2}\right) + \frac{\Lambda}{4}a^4  \nonumber\\
    & + 16 V_{\rm IX}(\beta_{+},\beta_{-}) a^2 \Bigg] \Psi_{\rm GR}(a,\beta_{+},\beta_{-}) = 0.
\end{align}
Here $\Psi_{\rm GR}(a,\beta_{+},\beta_{-})$ is the wave function of the Bianchi IX universe with a cosmological constant according to General Relativity. The parameter $p$ characterizes some of the operator ordering ambiguities \cite{Hawking:1985bk,PhysRevD.37.888,Kontoleon:1998pw,Mondal:2025qyd}. However, it turns out that the potential $V_{\rm IX}$ is too complicated, and assumptions must be made to get an analytical handle on the problem. In the small anisotropy limit, the potential simplifies to
\begin{align}
    V_{\rm IX}(\beta_{+},\beta_{-}) = - \frac{3}{64} + \frac{3}{8} \left( \beta_{+}^2 + \beta_{-}^2 \right) + \mathcal{O}(\beta_{\pm}^3,\beta_{\pm}\beta_{\mp}^2),
\end{align}
and the Wheeler-DeWitt equation takes the following form
\begin{align}\label{eq:WDW_GR}
    & \Bigg[ \hbar^2 \partial_{a}^2 - \frac{\hbar^2}{a^2} \frac{\boldsymbol{\partial}^2}{ \boldsymbol{\partial\beta}^2} + \frac{9\pi^4}{4\kappa^4}\Bigg(\frac{\Lambda}{3}a^4  - a^2 \nonumber\\
    & ~~~~~~ + 8 a^2 \boldsymbol{\beta}^2 \Bigg) \Bigg] \Psi_{\rm GR}(a,\boldsymbol{\beta}) = 0.
\end{align}
where we have used the notations
\begin{align}\label{eq:notation}
    \boldsymbol{\beta}^2 \equiv \beta_{+}^2 + \beta_{-}^2, ~~~ \frac{\boldsymbol{\partial}^2}{ \boldsymbol{\partial\beta}^2} \equiv \frac{\partial^2}{\partial \beta_{+}^2} + \frac{\partial^2}{\partial \beta_{-}^2}. 
\end{align}
Exact analytic solutions of the above Wheeler-DeWitt equation are not known, as the equation is not separable into isotropic and anisotropic parts. However, as anisotropy is treated as a perturbation, one can build on the solution $\psi_{\rm GR}^{(0)}(a)$ which solves the isotropic Wheeler-DeWitt equation and construct the perturbed Wheeler-DeWitt equation as
\begin{align}
    \Psi_{\rm GR}(a,\boldsymbol{\beta}) \simeq \psi^{(0)}_{\rm GR}(a) \, \chi_{\rm GR} (a,\boldsymbol{\beta}).
\end{align}
More specifically, we use the following WKB \textit{ansatz}
\begin{align}
    \Psi_{\rm GR} \simeq & \e^{\frac{1}{\hbar}\left(S_{0}(a) + \hbar S_{1}(a) + \frac{1}{2}\sigma(a) \boldsymbol{\beta}^2\right)}.
\end{align}
Plugging this ansatz into the Wheeler-DeWitt equation, we have to solve order-by-order in both $\hbar$ and $\boldsymbol{\beta}$, where in each order the Wheeler-DeWitt equation reads
\begin{subequations}\label{eq:wkb_equations}
\begin{align}
    & \hbar^0,\boldsymbol{\beta}^0:~~~ (S'_{0})^2 + \frac{9\pi^4}{4\kappa^4} \left(\frac{\Lambda}{3}a^4 - a^2 \right) = 0, \\
    & \hbar^0,\boldsymbol{\beta}^2:~~~ S'_{0} \sigma' - \frac{\sigma^2}{a^2} + \frac{18\pi^4}{\kappa^4} a^2 = 0, \label{eq:Riccati_GR} \\
    & \hbar^1,\boldsymbol{\beta}^0:~~~  S''_{0} + \frac{p}{a} S'_{0}  + 2 S'_{0} S'_{1} -2 \frac{\sigma}{a^2} = 0.
\end{align}    
\end{subequations}
Notice that the isotropic part of the superpotential has a turning point at $a = \sqrt{3/\Lambda}$. The solutions to the $\mathcal{O}(\hbar^0,\boldsymbol{\beta}^0)$ equation
\begin{align}
    S'_{0} = s_{1} \frac{3\pi^2}{2\kappa^2} \sqrt{a^2 - \frac{\Lambda}{3}a^4},
\end{align}
where $s_{1} = \pm 1$, on both sides of the turning point read as  
\begin{align}
    S_{0} = \begin{cases}
         s_{1}\frac{3\pi^2 i}{2\kappa^2\Lambda}\left(\frac{\Lambda}{3}a^2-1\right)^{\frac{3}{2}}, & a>\sqrt{\frac{3}{\Lambda}}, \\
        - s_{1} \frac{3\pi^2}{2\kappa^2\Lambda}\left(1-\frac{\Lambda}{3}a^2\right)^{\frac{3}{2}}, & a<\sqrt{\frac{3}{\Lambda}}.
    \end{cases}
\end{align}
In the classically accessible region $a>\sqrt{3/\Lambda}$, the momentum corresponding to the leading order wave function $\e^{\frac{i}{\hbar}S_{0}}$ is given by
\begin{align}
    - i \hbar \frac{\rd}{\rd a}\e^{\frac{S_{0}}{\hbar}} = \underbrace{s_{1} \frac{3\pi^2 a}{2\kappa^2} \left(\frac{\Lambda}{3}a^2 - 1\right)^{\frac{1}{2}}}_{{\rm momentum}} \e^{\frac{S_{0}}{\hbar}}.
\end{align}
Due to the unusual negative sign in the definition of conjugate momentum $p_{a}$ in \eqref{eq:conjugate_momenta}, the state with $s_{1}=+1$ corresponds to a contracting universe ($\dot{a}<0$), whereas the $s=-1$ state corresponds to an expanding universe ($\dot{a}>0$) in the classically accessible region. Note that the notion of the classically accessible/forbidden region only makes sense in the isotropic part of the wave function, since when the anisotropies are included, the differential operator becomes hyperbolic, and the notion of accessible/forbidden may not apply in many cases (recall Klein paradox in relativistic quantum mechanics in the context of Klein-Gordon equation).

In the next order $\mathcal{O}(\hbar^0,\boldsymbol{\beta}^2)$, we solve for the anisotropy function $\sigma(a)$. It can be shown that the Riccati equation \eqref{eq:Riccati_GR} has the following solutions
\begin{align}\label{eq:sigma}
    \sigma(a) = \begin{cases}
        \frac{-\frac{12\pi^2}{\kappa^2}a^2}{s_{1}i\sqrt{\frac{\Lambda}{3}a^2-1} +s_{2} 3}, & a>\sqrt{\frac{3}{\Lambda}}, \\
        \frac{-\frac{12\pi^2}{\kappa^2}a^2}{s_{1}\sqrt{1-\frac{\Lambda}{3}a^2} + s_{2} 3}, & a<\sqrt{\frac{3}{\Lambda}}.
    \end{cases}
\end{align}
where $s_{2} = \pm 1$. See \cite{deAlwis:2018sec} for a similar solution in the context of inhomogeneous perturbations. Notice that $\sigma(a)$ can also be expressed as
\begin{align}
    \sigma(a) = \begin{cases}
        \frac{-\frac{12\pi^2}{\kappa^2}a^2\left(-s_{1}i\sqrt{\frac{\Lambda}{3}a^2-1} + s_{2} 3\right)}{\frac{\Lambda}{3}a^2+8}, & a>\sqrt{\frac{3}{\Lambda}}, \\
        \frac{-\frac{12\pi^2}{\kappa^2}a^2\left(-s_{1}\sqrt{1-\frac{\Lambda}{3}a^2} + s_{2} 3\right)}{\frac{\Lambda}{3}a^2+8}, & a<\sqrt{\frac{3}{\Lambda}}.
    \end{cases}
\end{align}
Since $\frac{\Lambda}{3}a^2+8>0$, in both the regions $a\lessgtr\sqrt{3/\Lambda}$, we must have $s_{2}=+1$, for stability under anisotropic perturbations. When $s_{2}=-1$, the anisotropic wave function takes an inverse Gaussian form, favoring the growth of anisotropies. This choice has to be discarded on physical grounds. It is interesting to note that the regularity condition $\sigma(a)<0$ fixes the integration constant that the Riccati equation \eqref{eq:Riccati_GR} would otherwise allow for and makes the solution \eqref{eq:sigma} unique. This can be seen from the series solution of \eqref{eq:Riccati_GR} around $a=0$, where choosing the regular solution fixes all the subsequent series coefficients.

The solution for $\mathcal{O}(\hbar^1,\boldsymbol{\beta}^0)$ is given by
\begin{align}
    S_{1} = \begin{cases}
        \log\left(\frac{a^{-\frac{1}{2}(1+p)}}{\left(\frac{\Lambda}{3}a^2 - 1\right)^{\frac{1}{4}}}\frac{\left(i\sqrt{\frac{\Lambda}{3}a^2 - 1} + s_{1} \right)^2}{\left(\frac{\Lambda}{3} a^2 + 2 - 2i 
 s_{1} \sqrt{\frac{\Lambda}{3}a^2 - 1} \right)}\right),  & a>\sqrt{\frac{3}{\Lambda}} \\
        \log \left(\frac{a^{-\frac{1}{2}(1+p)}}{\left(1 - \frac{\Lambda}{3}a^2\right)^{\frac{1}{4}}} \frac{\left(\sqrt{1-\frac{\Lambda}{3}a^2} + s_{1} \right)^2}{\left(\frac{\Lambda}{3} a^2 + 2 - 2 s_{1} \sqrt{1-\frac{\Lambda}{3}a^2} \right)}\right) ,  & a<\sqrt{\frac{3}{\Lambda}} \\
    \end{cases}
\end{align}
Notice that the prefactor in the wave function $a^{-\frac{1}{2}(1+p)} \left(\frac{\Lambda}{3}a^2 - 1\right)^{-\frac{1}{4}}$ appears in the isotropic case as well, whereas the other remaining factor appears specifically due to the presence of anisotropy. It is also interesting to note that the operator ordering ambiguity parameter $p$ only starts to appear in $\mathcal{O}(\hbar)$ and is absent in the semi-classical wave function at $\mathcal{O}(\hbar^0)$, which means that at the leading order, the ambiguity is irrelevant.

Then, the most general solution can be written as a linear combination of expanding and contracting branches in the classically accessible region and exponentially growing and suppressing branches in the classically forbidden region, along with their associated anisotropic counterparts
\begin{widetext}
\begin{align}
    \Psi_{\rm GR}(a,\boldsymbol{\beta}) \simeq
\begin{cases}
A\frac{a^{-\frac{1}{2}(1+p)}}{\left(\frac{\Lambda}{3}a^2 - 1\right)^{\frac{1}{4}}}\frac{\left(i\sqrt{\frac{\Lambda}{3}a^2 - 1} + 1 \right)^2}{\left(\frac{\Lambda}{3} a^2 + 2 - 2i 
\sqrt{\frac{\Lambda}{3}a^2 - 1} \right)} \e^{+i\frac{3\pi^2}{2\kappa^2\Lambda\hbar}\left(\frac{\Lambda}{3}a^2-1\right)^{\frac{3}{2}} - \frac{\frac{6\pi^2}{\kappa^2 \hbar}a^2}{i\sqrt{\frac{\Lambda}{3}a^2-1} + 3} \boldsymbol{\beta}^2} \\
 ~~~~~~~~~~~~+B \frac{a^{-\frac{1}{2}(1+p)}}{\left(\frac{\Lambda}{3}a^2 - 1\right)^{\frac{1}{4}}}\frac{\left(i\sqrt{\frac{\Lambda}{3}a^2 - 1} - 1 \right)^2}{\left(\frac{\Lambda}{3} a^2 + 2 + 2i \sqrt{\frac{\Lambda}{3}a^2 - 1} \right)} \e^{-i\frac{3\pi^2}{2\kappa^2\Lambda\hbar}\left(\frac{\Lambda}{3}a^2-1\right)^{\frac{3}{2}} - \frac{\frac{6\pi^2}{\kappa^2\hbar}a^2}{-i\sqrt{\frac{\Lambda}{3}a^2-1} + 3} \boldsymbol{\beta}^2}, & \text{for, } a> \sqrt{\frac{3}{\Lambda}}, \\
C \frac{a^{-\frac{1}{2}(1+p)}}{\left(1 - \frac{\Lambda}{3}a^2\right)^{\frac{1}{4}}} \frac{\left(\sqrt{1-\frac{\Lambda}{3}a^2} + 1 \right)^2}{\left(\frac{\Lambda}{3} a^2 + 2 - 2 \sqrt{1-\frac{\Lambda}{3}a^2} \right)} \e^{- \frac{3\pi^2}{2\kappa^2\Lambda\hbar}\left(1-\frac{\Lambda}{3}a^2\right)^{\frac{3}{2}} - \frac{+\frac{6\pi^2}{\kappa^2\hbar}a^2}{\sqrt{1-\frac{\Lambda}{3}a^2} + 3}\boldsymbol{\beta}^2} \\
~~~~~~~~~~~~+ D \frac{a^{-\frac{1}{2}(1+p)}}{\left(1 - \frac{\Lambda}{3}a^2\right)^{\frac{1}{4}}} \frac{\left(\sqrt{1-\frac{\Lambda}{3}a^2} - 1 \right)^2}{\left(\frac{\Lambda}{3} a^2 + 2 + 2 \sqrt{1-\frac{\Lambda}{3}a^2} \right)} \e^{+ \frac{3\pi^2}{2\kappa^2\Lambda\hbar}\left(1-\frac{\Lambda}{3}a^2\right)^{\frac{3}{2}} - \frac{+\frac{6\pi^2}{\kappa^2\hbar}a^2}{-\sqrt{1-\frac{\Lambda}{3}a^2} + 3}\boldsymbol{\beta}^2}, & \text{for, } a< \sqrt{\frac{3}{\Lambda}}.
\end{cases}
\end{align}    
\end{widetext}
where $A,B,C,$ and $D$ are constants. Notice that the isotropic part of the Wheeler-DeWitt equation is a second-order differential equation leading to two unknown constants. On the other hand, the anisotropic wave function is uniquely determined by the regularity condition or by the choice $s_{2}=+1$. Thus, among the four constants, only two are independent, and the other two have to be determined by using the WKB matching conditions across the turning point. The two independent constants have to be determined by physical arguments. In this regard, for the isotropic theory, there are several proposals, most famous among these being the Hartle-Hawking no-boundary proposal and Vilenkin's tunneling proposal.

In the tunneling proposal, based on the current observation, only the expanding branch of the universe is selected in the classically accessible region, \textit{i.e.}, $A=0$. Then the solution in the allowed region can be analytically continued to the classically forbidden region along a path in the complex plane, avoiding the turning point where the wave function diverges. Among the possible branches $(\frac{\Lambda}{3}a^2-1)\to (1-\frac{\Lambda}{3}a^2)\e^{\pm i \pi}$, choosing the branch $\e^{-i\pi}$ allows us to identify
\begin{align}
    B \e^{+i\frac{\pi}{4}} = D.
\end{align}
Also, analytically continuing along the other branch $\e^{+i\pi}$ produces the exponentially suppressed term, which is subdominant away from the turning point, and we discard this term by setting $C=0$.

To determine the correct coefficient for the subdominant piece, one should carefully define the path in the complex plane and study the Stokes phenomenon \cite{Berry:1972na}, which is a fairly involved discussion in itself. However, here we are only interested in approximate expressions for the wave function, which are valid up to small corrections from subdominant pieces.

On the other hand, the Hartle-Hawking wave function is motivated by Euclidean quantum gravity. In this proposal near $a\sim 0$, the branch with negative imaginary momentum is chosen \cite{di2019noprescription}, which corresponds to the choice $D=0$. In this case, analytically continuing from the classically forbidden to the allowed region, $(1-\frac{\Lambda}{3}a^2)\to (\frac{\Lambda}{3}a^2-1)\e^{\pm i \pi}$, we identify
\begin{align}
    C\e^{-i\frac{\pi}{4}} = A, ~~~ C\e^{+i\frac{\pi}{4}} = B
\end{align}
Thus, the tunneling and Hartle-Hawking wave functions are given respectively by
\begin{widetext}
\begin{align}
    \Psi_{\rm GR}^{\rm (T)}(a,\boldsymbol{\beta}) \simeq
\begin{cases}
B \frac{a^{-\frac{1}{2}(1+p)}}{\left(\frac{\Lambda}{3}a^2 - 1\right)^{\frac{1}{4}}}\frac{\left(i\sqrt{\frac{\Lambda}{3}a^2 - 1} - 1 \right)^2}{\left(\frac{\Lambda}{3} a^2 + 2 + 2i \sqrt{\frac{\Lambda}{3}a^2 - 1} \right)} \e^{-i\frac{3\pi^2}{2\kappa^2\Lambda\hbar}\left(\frac{\Lambda}{3}a^2-1\right)^{\frac{3}{2}} - \frac{\frac{6\pi^2}{\kappa^2\hbar}a^2}{-i\sqrt{\frac{\Lambda}{3}a^2-1} + 3} \boldsymbol{\beta}^2}, & \text{for, } a> \sqrt{\frac{3}{\Lambda}}, \\
B \e^{i\frac{\pi}{4}} \frac{a^{-\frac{1}{2}(1+p)}}{\left(1 - \frac{\Lambda}{3}a^2\right)^{\frac{1}{4}}} \frac{\left(\sqrt{1-\frac{\Lambda}{3}a^2} - 1 \right)^2}{\left(\frac{\Lambda}{3} a^2 + 2 + 2 \sqrt{1-\frac{\Lambda}{3}a^2} \right)} \e^{+ \frac{3\pi^2}{2\kappa^2\Lambda\hbar}\left(1-\frac{\Lambda}{3}a^2\right)^{\frac{3}{2}} - \frac{+\frac{6\pi^2}{\kappa^2\hbar}a^2}{-\sqrt{1-\frac{\Lambda}{3}a^2} + 3}\boldsymbol{\beta}^2}, & \text{for, } a< \sqrt{\frac{3}{\Lambda}},
\end{cases}\label{eq:tunneling_wave_function}
\end{align}
and 
\begin{align}\label{eq:HH_anisotropic}
    \Psi_{\rm GR}^{\rm (HH)}(a,\boldsymbol{\beta}) \simeq
\begin{cases}
C\e^{-i\frac{\pi}{4}}\frac{a^{-\frac{1}{2}(1+p)}}{\left(\frac{\Lambda}{3}a^2 - 1\right)^{\frac{1}{4}}}\frac{\left(i\sqrt{\frac{\Lambda}{3}a^2 - 1} + 1 \right)^2}{\left(\frac{\Lambda}{3} a^2 + 2 - 2i 
\sqrt{\frac{\Lambda}{3}a^2 - 1} \right)} \e^{+i\frac{3\pi^2}{2\kappa^2\Lambda\hbar}\left(\frac{\Lambda}{3}a^2-1\right)^{\frac{3}{2}} - \frac{\frac{6\pi^2}{\kappa^2 \hbar}a^2}{i\sqrt{\frac{\Lambda}{3}a^2-1} + 3} \boldsymbol{\beta}^2} \\
 ~~~~~~~~~~~~+C\e^{+i\frac{\pi}{4}} \frac{a^{-\frac{1}{2}(1+p)}}{\left(\frac{\Lambda}{3}a^2 - 1\right)^{\frac{1}{4}}}\frac{\left(i\sqrt{\frac{\Lambda}{3}a^2 - 1} - 1 \right)^2}{\left(\frac{\Lambda}{3} a^2 + 2 + 2i \sqrt{\frac{\Lambda}{3}a^2 - 1} \right)} \e^{-i\frac{3\pi^2}{2\kappa^2\Lambda\hbar}\left(\frac{\Lambda}{3}a^2-1\right)^{\frac{3}{2}} - \frac{\frac{6\pi^2}{\kappa^2\hbar}a^2}{-i\sqrt{\frac{\Lambda}{3}a^2-1} + 3} \boldsymbol{\beta}^2}, & \text{for, } a> \sqrt{\frac{3}{\Lambda}}, \\
C \frac{a^{-\frac{1}{2}(1+p)}}{\left(1 - \frac{\Lambda}{3}a^2\right)^{\frac{1}{4}}} \frac{\left(\sqrt{1-\frac{\Lambda}{3}a^2} + 1 \right)^2}{\left(\frac{\Lambda}{3} a^2 + 2 - 2 \sqrt{1-\frac{\Lambda}{3}a^2} \right)} \e^{- \frac{3\pi^2}{2\kappa^2\Lambda\hbar}\left(1-\frac{\Lambda}{3}a^2\right)^{\frac{3}{2}} - \frac{+\frac{6\pi^2}{\kappa^2\hbar}a^2}{\sqrt{1-\frac{\Lambda}{3}a^2} + 3}\boldsymbol{\beta}^2}, & \text{for, } a< \sqrt{\frac{3}{\Lambda}}.
\end{cases}
\end{align}
\end{widetext}
Notice that the expression for the tunneling wave function \eqref{eq:tunneling_wave_function} matches exactly that in \cite{delCampo:1989hy}. We note that the authors of \cite{delCampo:1989hy} also have ignored the subdominant piece, as we have done here.

 Therefore, we have determined the wave function of an anisotropic universe in General Relativity up to an overall normalization constant. Having discussed the General Relativity case, we now turn to the Ho\v{r}ava-Lifshitz theory. We will see that the wave function in the Ho\v{r}ava-Lifshitz theory has correspondence with the Hartle-Hawking wave function in the large universe $(a\gg1)$ limit.

\section{Wave function of an anisotropic universe in Ho\v{r}ava-Lifshitz gravity}\label{sec:HL_wave_function}
In the Horava-Lifshitz theory, an anisotropic scaling between the space and time, $(t,\boldsymbol{x}) \to (b^{z} t,b\boldsymbol{x})$ is proposed in the UV regime. Such an anisotropic scaling is also known as the Lifshitz scaling and is responsible for the UV violation of Lorentz symmetry. for the case of the critical exponent $z=3$, the Ho\v{r}ava-Lifshitz theory has been shown to be perturbatively renormalizable \cite{Barvinsky:2015kil}. The fundamental symmetry of the theory is the foliation-preserving diffeomorphisms: space-independent time reparametrizations and time-dependent spatial diffeomorphisms of the spatial hypersurfaces. In the projectable version of the Ho\v{r}ava-Lifshitz gravity the lapse function is assumed to be a function of time only $N=N(t)$. This is known as the \textit{projectability} condition. The action for the projectable Ho\v{r}ava-Lifshitz gravity consistent with these symmetries and condition can be written as
\begin{align}
    \mathscr{A} = \frac{1}{2\kappa^2} \int \rd t \rd^3 x \sqrt{h} N \left(\mathscr{K}_{ij}\mathscr{K}^{ij} - \lambda \mathscr{K}^2 - \mathscr{V}_{\rm HL}\right),
\end{align}
where the Ho\v rava-Lifshitz potential $\mathscr{V}_{\rm HL}$ containing higher order spatial derivatives is given by
\begin{align}
    \mathscr{V}_{\rm HL} = & 2 \Lambda + g_{1} \mathscr{R} + \kappa^2 \left(g_{2} \mathscr{R}^2 + g_{3} \mathscr{R}^i_j \mathscr{R}_i^j\right)  \nonumber\\
    & + \kappa^4 \Big(g_{5}\mathscr{R}^3+g_{6}\mathscr{R}\mathscr{R}^i_j\mathscr{R}^j_i+g_{7}\mathscr{R}^{i}_{j} \mathscr{R}^{j}_{k} \mathscr{R}^{k}_{i} \nonumber\\
    & +g_{8}\mathscr{R} {}^{(3)}\Delta \mathscr{R}+g_{9} {}^{(3)}\nabla_{i} \mathscr{R}_{jk} {}^{(3)}\nabla^{i} \mathscr{R}^{jk}\Big),
\end{align}
where $\mathscr{R}_{ij}$ is the 3-Ricci tensor, ${}^{(3)}\nabla_{i}$ is the spatial covariant derivative and ${}^{(3)}\Delta\equiv {}^{(3)}\nabla^{i}{}^{(3)}\nabla_{i}$ is the 3-Laplacian. The coupling parameters $\lambda$ and $g_{i}$s should run according to the renormalization group (RG) flow. Notice that we have avoided adding the CPT violating term $\kappa^3 g_{4}  \epsilon^{ijk} \mathscr{R}_{il} \nabla_{j} \mathscr{R}^l_k$ since CPT violation has stringent observational constraint and as a result it is reasonable to assume that quantum gravity should be CPT preserving \cite{Toma:2012xa}.  The constant $g_{1}$ can be set to $-1$. The critical exponent is expected to flow from $z=3$ in the UV regime to $z=1$ in the IR regime. If by means of RG flow $\lambda$ acquires the value $1$ in the IR regime, then one potentially finds agreement with General Relativity in this regime.

Due to the fact that the lapse function is only a function of time, the variation with respect to lapse leads to a Hamiltonian constraint that is not local but integrated over the whole space. However, as we are working with a homogeneous universe and we assumed for simplicity that the space consists of only one connected piece, the integral is trivial and results in a factor of co-moving 3-volume, which is in this case $2\pi^2$. Therefore, we have a Hamiltonian constraint, which can be promoted to an appropriate operator leading to the Wheeler-DeWitt equation for a homogeneous, anisotropic universe corresponding to the Ho\v{r}ava-Lifshitz theory. However, extending the theory to include inhomogeneities or/and to consider the space consisting of more than one connected pieces leads to the ``dark matter as integration constant''~\cite{Mukohyama:2009mz}. This would introduce additional separation constants in the study of the Wheeler-DeWitt equation, see for example \cite{Matsui:2021yte}.

For the Bianchi IX metric \ref{eq:bianchi_ix}, $\mathscr{V}_{\rm HL}$ takes the form
\begin{align}
     \mathscr{V}_{\rm HL}= & \frac{128}{a^6} \Bigg(\frac{\Lambda}{64}a^6 + V_{4} a^4 + \kappa^2 V_{2} a^2 + \kappa^4 V_{0} \Bigg),
\end{align}
where the quantities $V_{0,2,4}$  are given by
\begin{widetext}
\begin{align}
    V_{4} = & \frac{g_{1}}{16} \left(\frac{1}{2}\e^{4\beta_{+}}\left(1-\cosh 4\sqrt{3}\beta_{-}\right)+\e^{-2\beta_{+}}\cosh 2\sqrt{3}\beta_{-}-\frac{1}{4}\e^{-8\beta_{+}}\right),\\
    V_{2} = & \frac{\e^{8\beta_{+}}}{16}
    \left(3 g_{2} + g_{3}
    -4 (g_{2}+g_{3})\cosh 4\sqrt{3}\beta_{-}+(g_{2}+3g_{3})\cosh 8\sqrt{3}\beta_{-}\right)+  (g_{2}+g_{3}) \frac{\e^{2\beta_{+}}}{4}  \left(\cosh 2\sqrt{3}\beta_{-} - \cosh 6\sqrt{3}\beta_{-} \right) \nonumber\\
    & +\frac{1}{8}\e^{-4\beta_{+}}\left((3g_{2}+g_{3})\cosh 4\sqrt{3}\beta_{-} + g_{2} + g_{3} \right)  - \frac{1}{4} (g_{2}+g_{3}) \e^{-10\beta_{-}} \cosh 2\sqrt{3}\beta_{-} +\frac{1}{32}(g_{2}+3g_{3})\e^{-16\beta_{+}},\\
    V_{0} = & \e^{12 \beta_{+}}\Bigg(-\frac{1}{8}\left(g_{5} + 3 g_{6}+g_{7} -8 g_{9} \right)\cosh 12 \sqrt{3} \beta_{-}-\frac{1}{8}\left(15g_{5}+13g_{6}+15g_{7}-24g_{9} \right)\cosh 4\sqrt{3}\beta_{-}\nonumber\\
    & + \frac{1}{4}\left(3g_{5} + 5 g_{6} + 3g_{7} -4g_{9}\right)\cosh 8\sqrt{3}\beta_{-}+\frac{1}{4}\left(5g_{5}+3g_{6}+5g_{7}-12g_{9}\right)\Bigg)\nonumber\\
   & + \e^{6 \beta_{+}} \Bigg(\frac{1}{4} (3 g_{5} + 5 g_{6}+3g_{7}-4g_{9})\cosh 10 \sqrt{3} \beta_{-} - \frac{1}{4} (9g_{5}+7g_{6}-3g_{7}+20g_{9})
  \cosh 6 \sqrt{3} \beta_{-} \nonumber\\
  & + \frac{1}{2} (3g_{5}+g_{6}-3g_{7}+12g_{9}) \cosh 2 \sqrt{3}
   \beta_{-}\Bigg) + \frac{1}{2} (3g_{5}+g_{6}-3g_{7}+12g_{9}) \cosh{4 \sqrt{3} \beta_{-}}   \nonumber\\
   &  - \frac{1}{8} (15g_{5}+13g_{6}+15g_{7}-24g_{9}) \cosh{8 \sqrt{3} \beta_{-}}  +\frac{1}{8}(3g_{5}+9g_{6}+27g_{7}-72g_{9}) \nonumber\\
   & +\e^{-6 \beta_{+}} \left( \frac{1}{2} (5g_{5}+3g_{6}+5g_{7}-12g_{9})\cosh 6 \sqrt{3} \beta_{-} + \frac{1}{2} (3g_{5}+g_{6}-3g_{7}+12g_{9}) \cosh 2 \sqrt{3} \beta_{-}\right) \nonumber\\
   & +\e^{-12 \beta_{+}} \left(-\frac{1}{8}(15g_{5}+13g_{6}+15g_{7}-24g_{9})\cosh 4
   \sqrt{3} \beta_{-} - \frac{1}{8} (9g_{5}+7g_{6}-3g_{7}+20g_{9})\right)   \nonumber\\
   &  +\e^{-18 \beta_{+}} \left(\frac{1}{4}(3g_{5}+5g_{6}+3g_{7}-4g_{9})\cosh 2 \sqrt{3} \beta_{-}\right) - \frac{1}{16}(g_{5} + 3g_{6}+g_{7}-8g_{9})\e^{
   -24 \beta_{+}}.
\end{align}
\end{widetext}
For the study of classical dynamics of Bianchi IX spacetime in Ho\v{r}ava-Lifshitz gravity, where a potential similar to the above $\mathscr{V}_{\rm HL}$ appears, see \cite{Bakas:2009ku,Bakas:2010fm,Misonoh:2011mn}.

Thus, the action for the Bianchi IX universe according to the Ho\v{r}ava-Lifshitz theory is
\begin{align}\label{eq:horava_lifshitz_action}
    \mathscr{A} = & \frac{\pi^2}{8\kappa^2} \int \rd t  \Bigg[-\frac{3(3\lambda-1)}{N} a \dot{a}^2 + \frac{6}{N}a^3 (\dot{\beta}_{+}^2 + \dot{\beta}_{-}^2) \nonumber\\
    & - N \frac{128}{a^3} \Bigg(\frac{\Lambda}{64}a^6 + V_{4} a^4 + \kappa^2 V_{2} a^2 + \kappa^4 V_{0} \Bigg)\Bigg].
\end{align}
Apart from the additional terms in the potential, the kinetic term in Ho\v{r}ava-Lifshitz action is also non-perturbatively different from that in the Einstein-Hilbert action \eqref{eq:einstein-hilbert} due to the parameter $\lambda$. For $\lambda=1$, the kinetic term coincides with that in \eqref{eq:einstein-hilbert}. The Hamiltonian corresponding to the action \eqref{eq:horava_lifshitz_action} is given by
\begin{align}
    H \mathscr{H} = & N\left[ - \frac{2\kappa^2}{3\pi^2(3\lambda-1)} \frac{p_{a}^2}{a} + \frac{\kappa^2}{3\pi^2a^3} \left(p_{\beta_{+}}^2 + p_{\beta_{-}}^2\right) \right. \nonumber\\
    & \left. + \frac{16\pi^2}{\kappa^2a^3} \Bigg(\frac{\Lambda}{64}a^6 + V_{4} a^4 + \kappa^2 V_{2} a^2 + \kappa^4 V_{0} \Bigg) \right],
\end{align}
where the conjugate momenta are defined as
\begin{align}\label{eq:conjugate_momenta_HL}
    p_{a} = -\frac{\pi^2}{4\kappa^2} \frac{3(3\lambda-1)}{N} a \dot{a},~~~~ p_{\beta_{\pm}} = \frac{3\pi^2}{2N \kappa^2} a^3 \dot{\beta}_{\pm}.
\end{align}
We can again write the Wheeler-DeWitt equation by promoting the variables to operators as follows
\begin{align}\label{eq:WDW_HL}
    & \left[\frac{2\kappa^2 \hbar^2}{3\pi^2(3\lambda-1)} a^{-p}\frac{\partial}{\partial a} a^{p}\frac{\partial}{\partial a} - \frac{\kappa^2 \hbar^2}{3\pi^2a^2} \left(\frac{\partial^2}{\partial \beta_{+}^2} + \frac{\partial^2}{\partial \beta_{-}^2} \right) \right. \nonumber \\
    & \left. + \frac{16\pi^2}{\kappa^2} \Bigg(\frac{\Lambda}{64}a^4 + V_{4} a^2 + \kappa^2 V_{2} + \frac{\kappa^4}{a^2} V_{0} \Bigg) \right] \Psi_{\rm HL} = 0.
\end{align}
Here, the parameter $p$ again characterizes some of the operator ordering ambiguities. As the anisotropic potential is too complicated, the system cannot be tackled analytically without simplifying assumptions. Like in the General Relativity case, we assume that a closed, homogeneous, and isotropic universe is deformed to introduce a small amount of anisotropy. In this small anisotropy limit, the various terms in the anisotropy potential take the following form
\begin{subequations}
    \begin{align}
        V_{0} & = \frac{3}{16}(9g_{5}+3g_{6}+g_{7})-\frac{9}{2}(9g_{5}-g_{6}-3g_{7}-4g_{9}) \nonumber\\
        & \qquad\qquad \qquad \times \left( \beta_{+}^2 + \beta_{-}^2 \right)  \\
        V_{2} & = \frac{3}{32}(3g_{2}+g_{3}) - \frac{3}{2} (3g_{2}-g_{3}) \left( \beta_{+}^2 + \beta_{-}^2 \right)\\
        V_{4} & = - \frac{3}{64} + \frac{3}{8} \left( \beta_{+}^2 + \beta_{-}^2 \right)
    \end{align}
\end{subequations}
Here $V_{4}$ is exactly the same as $V_{\rm IX}$ in the General Relativity case. $V_{0}$ and $V_{2}$ are the terms due to Ho\v{r}ava-Lifshitz theory. Like before, we now try to solve the Wheeler-DeWitt equation with the WKB approximation, wherein the wave function of the universe is approximately separated as
\begin{align}
    \Psi_{\rm HL}(a,\boldsymbol{\beta}) \simeq \psi^{(0)}_{\rm HL} \, \chi_{\rm HL} (a,\boldsymbol{\beta}).
\end{align}
The function $\psi^{(0)}_{\rm HL}$ solves the Wheeler-DeWitt equation in the isotropic limit, and the function $\chi_{\rm HL} (a,\boldsymbol{\beta})$ is determined by solving the Wheeler-DeWitt equation in a power series on both $\hbar$ and anisotropy magnitude $(\beta_{+}^2 + \beta_{-}^2)$. In the following, we describe the procedure in more details.

\begin{figure}
    \centering
    \includegraphics[width=\linewidth]{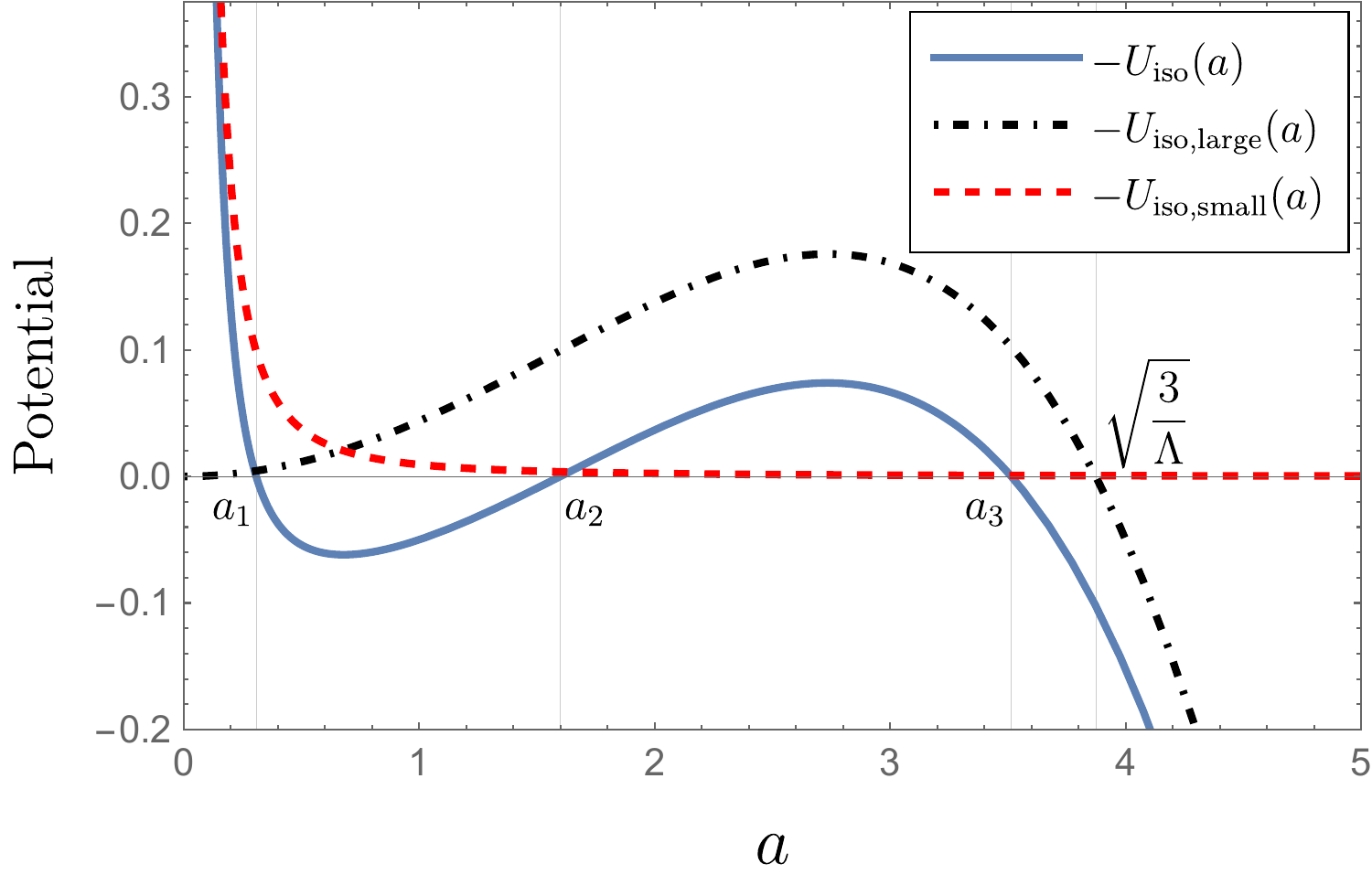}
    \caption{We consider a scenario where the isotropic potential $-U_{\rm iso}(a)$ given in \eqref{eq:complete_isotropic_potential_HL} has three turning points denoted as $a_{1,2,3}$. Approximate expressions for the potential in the large universe limit ($a\gg1$) denoted as $-U_{\rm iso,large}(a)$ and in the small universe limit ($a\ll 1$) denoted as $-U_{\rm iso,small}(a)$ are given in \eqref{eq:large_potential_iso_HL} and \eqref{eq:potential_small_HL}, respectively. These approximate expressions do not capture the structure of the Ho\v{r}ava-Lifshitz potential $-U_{\rm iso}(a)$ in the intermediate region.}
    \label{fig:UV_IR_Extreme}
\end{figure}

We rewrite the Wheeler-DeWitt equation \eqref{eq:WDW_HL} in the following form 
\begin{align}\label{eq:WDW_HL_1}
    & \left[\hbar^2 \left(\frac{\partial^2}{\partial a^2} + \frac{p}{a}\frac{\partial}{\partial a}\right) - \frac{\hbar^2}{ a^2} \left(\frac{3\lambda-1}{2}\right) \left(\frac{\boldsymbol{\partial}^2}{\boldsymbol{\partial \beta}^2} \right) \right. \nonumber \\
    & \left. + \frac{48\pi^4}{\kappa^4} \left(\frac{3\lambda-1}{2}\right)\left(U_{\rm iso}(a)+U_{\rm ani}(a) \boldsymbol{\beta}^2  \right) \right] \Psi_{\rm HL}(a,\boldsymbol{\beta}) = 0,
\end{align}
where we have again used the notation introduced in \eqref{eq:notation}, and the isotropic and anisotropic parts of the potential are defined as 
\begin{subequations}\label{eq:full_potential_2}
\begin{align}
     U_{\rm iso}(a) = & \frac{\Lambda}{64}a^4 - \frac{3}{64}a^2 +\frac{3\kappa^2}{32}g_{\rm r}+\frac{3\kappa^4}{16a^2} g_{\rm s}, \label{eq:complete_isotropic_potential_HL}\\
    U_{\rm ani} (a) = & \frac{3}{8}a^2 - \frac{3}{2}\kappa^2g_{A} -\frac{9\kappa^4}{2a^2} g_{B}.
\end{align}    
\end{subequations}
Here, for convenience, we have defined
\begin{equation}
    \begin{gathered}
     g_{\rm r} = (3g_{2}+g_{3}),~~~~~~ g_{\rm s}= (9g_{5}+3g_{6}+g_{7}), \\
    g_{A} = (3g_{2}-g_{3}),~~~~~~  g_{B}= (9g_{5}-g_{6}-3g_{7}-4g_{9}).   
    \end{gathered}
\end{equation}
We shall assume for the following discussions that we have a positive cosmological constant $\Lambda>0$. Furthermore, in the present paper we are interested in the case where $U_{\rm iso}(a)$ has three distinct, real and positive roots ($a_1<a_2<a_3$), as shown in \ref{fig:UV_IR_Extreme}. In this case we can consider emergence of an expanding universe with $a>a_3$ from an initial state of the universe trapped in the region $a_1<a<a_2$

\subsection{WKB \textit{ansatz} for the wave function}\label{sec:HL_wkb_setup}
Now, as before, let us take the following WKB \textit{ansatz} for the wave function
\begin{align}
    \Psi_{\rm HL} \sim & \e^{\frac{1}{\hbar}\left(S_{0}(a) + \hbar S_{1}(a) + \sqrt{\frac{2}{3\lambda-1}}\frac{2\sqrt{3}\pi^2}{\kappa^2} \Omega(a) \boldsymbol{\beta}^2\right)}.
\end{align}
Plugging this \textit{ansatz} into the Wheeler-DeWitt equation \eqref{eq:WDW_HL_1}, we obtain order-by-order the following set of equations
\begin{subequations}\label{eq:wkb_equations_hl}
\begin{align}
    & \hbar^0,\boldsymbol{\beta}^0:~~~ (S'_{0})^2 + \left(\frac{3\lambda-1}{2}\right) \frac{48\pi^4}{\kappa^4} U_{\rm iso}(a) = 0, \\
    & \hbar^0,\boldsymbol{\beta}^2:~~~ \frac{6\kappa^2}{\pi^2\sqrt{3\lambda-1}} S'_{0} \Omega' - 12 \frac{\Omega^2}{a^2} \nonumber\\
    & ~~~~~~~~~~~~~~ +12\left(\frac{3\lambda-1}{2}\right) U_{\rm ani}(a) = 0, \\
    & \hbar^1,\boldsymbol{\beta}^0:~~~  S''_{0} + \frac{p}{a} S'_{0}  + 2 S'_{0} S'_{1} \nonumber\\
    & ~~~~~~~~~~~~~~ -\frac{4\sqrt{6}\pi^2}{\kappa^2}\sqrt{3\lambda-1} \frac{\Omega}{a^2} = 0. \label{eq:first_order_wkb_hl}
\end{align}    
\end{subequations}
In general, the solution to the zeroth-order equation is
\begin{align}\label{eq:leading}
    S_{0}(a) = s_{1}\frac{4\sqrt{3}\pi^2}{\kappa^2} \sqrt{\frac{3\lambda-1}{2}}  \int^a \sqrt{- U_{\rm iso}(\tilde{a})} ~ \rd \tilde{a}.
\end{align}
where like before $s_{1}=\pm 1$ appears due to taking the square root. Then putting the above solution into the next order ($\mathcal{O}(\hbar^0,\boldsymbol{\beta}^2)$) results in the following differential equation for $\Omega$:
\begin{align}\label{eq:riccati}
    s_{1}\sqrt{-U_{\rm iso}(a)} ~ \Omega'(a)-\frac{\Omega^2(a)}{a^2} + \left(\frac{3\lambda-1}{2}\right) U_{\rm ani}(a) = 0.
\end{align}
This is again a Riccati equation. Analytical solutions for $S_{0}$, and $\Omega$ are available only for simpler instances of the potential. Both of these solutions are necessary to go to the next order, which we will discuss later. Let us first consider limiting cases: (a) the large $(a\gg1)$ and (b) the small universe $(a\ll 1)$ limit.

\subsection{Small universe \texorpdfstring{$(a\ll1)$}{} limit}\label{sec:wkb_small}

In this limit, the isotropic and anisotropic parts of the potential take the following simple forms
\begin{subequations}\label{eq:small_iso_aniso_HL}
 \begin{align}
    U_{\rm iso,small}(a) & \approx \frac{3\kappa^4}{16a^2} g_{\rm s} \label{eq:potential_small_HL}\\
    U_{\rm ani,small}(a) & \approx - \frac{9\kappa^4}{2a^2}g_{B}
\end{align}   
\end{subequations}
Under this assumption, from \eqref{eq:leading}, we can readily find $S_{0}$ by performing the integration and the result is
\begin{align}\label{eq:leading_wkb_small_hl}
    S_{0}(a) = s_{1}3\pi^2 \sqrt{\frac{3\lambda-1}{2}} \sqrt{-g_{\rm s}}  \log(a)+{\rm const.}
\end{align}
Moreover, notice that the Wheeler-DeWitt equation is separable into isotropic and anisotropic parts in the limit $a\ll1$. Thus, in this limit, we should look for a constant solution of $\Omega(a)$, which from \eqref{eq:riccati} turns out to be
\begin{align} \label{eq:constant_omega_hl}
    \Omega(a) = s_{2}\frac{3\kappa^2}{\sqrt{2}}\sqrt{\frac{3\lambda-1}{2}}\sqrt{-g_{B}},
\end{align}
where $s_{2}=\pm1$. For a stable or normalizable wave function for the anisotropies, we need
\begin{align}
    g_{B}<0,~~~ s_{2}=-1
\end{align}
The first inequality can be considered as a quantum gravitational bound on the coupling parameters of the Ho\v{r}ava-Lifshitz theory. We can now look for the solution for the next order $\mathcal{O}(\hbar^1,\boldsymbol{\beta}^0)$, and from \eqref{eq:first_order_wkb_hl}, using \eqref{eq:leading_wkb_small_hl} and \eqref{eq:constant_omega_hl}, we get
\begin{align}
    S_{1}(a) =& \frac{1}{2}\left(1-p-s_{1}\frac{4\sqrt{3(3\lambda-1)}\sqrt{-g_{B}}}{\sqrt{-g_{\rm s}}}\right) \log(a) \nonumber\\
    &+{\rm const.} 
\end{align}
Thus, the WKB solution in the small universe limit has the following form
\begin{align}
    \Psi_{\rm HL}^{\rm WKB}(a\ll 1) \simeq d_{+}a^{\delta_{+}} \exp\left(-\frac{3\kappa^2}{\hbar\sqrt{2}}\sqrt{\frac{3\lambda-1}{2}}\sqrt{-g_{B}}\boldsymbol{\beta}^2\right),
\end{align}
where, $d_{+}$ is a normalization constant and $\delta_{+}$ is given by
\begin{align}
    \delta_{+} =& \frac{1}{2}\left(1-p - \frac{4\sqrt{3(3\lambda-1)}\sqrt{-g_{B}}}{\sqrt{-g_{\rm s}}}\right)\nonumber\\
    & + \frac{3\pi^2}{\hbar} \sqrt{\frac{3\lambda-1}{2}} \sqrt{-g_{\rm s}}.
\end{align}
Here we have only considered the $s_{1}=+1$ solution due to the fact that in our numerical analysis of the Riccati equation \eqref{eq:riccati} for the complete potential \eqref{eq:full_potential_2} we found that only the $s_{1}=+1$ solutions lead to stable wave function for anisotropies under potential barriers. We will discuss this issue further in the section \ref{sec:numerical}. For a small universe, we see from \eqref{eq:potential_small_HL} that for $g_{\rm s}<0$, the universe lives under the potential barrier. The condition $g_{\rm s}<0$ also renders the wave function real. Further, notice that for the WKB analysis to be valid, one must also have $S_{0}/\hbar>S_{1}$, which implies
\begin{align}
    \left|\frac{3\pi^2}{\hbar} \sqrt{\frac{3\lambda-1}{2}} \sqrt{-g_{\rm s}} \right| > \left|\frac{1-p}{2} - \frac{2\sqrt{3(3\lambda-1)}\sqrt{-g_{B}}}{\sqrt{-g_{\rm s}}}\right|.
\end{align}
Thus, the WKB analysis in this region is valid for certain choices of the parameter values satisfying the above inequality. We may also demand $\delta_{+}>0$ in order for the wave function not to diverge in the limit $a\to0$. However, we note that in the $a\to0$ limit, the WKB approximation breaks down, and the above solution should not be trusted in this limit. We shall give further support to this point in our numerical analysis.

It is, however, not necessary to deploy WKB-type approximation in the small universe limit, and in this limit, the Wheeler-DeWitt equation separates into an isotropic and an anisotropic part, both of which are exactly solvable. We will discuss this solution in the section \ref{sec:small_exact_HL}.

\subsection{Large universe \texorpdfstring{$(a\gg1)$}{} limit}\label{sec:wkb_hl_large}
In the large universe limit $a\gg1$, the potential \eqref{eq:full_potential_2} assumes the following approximate form
\begin{subequations}\label{eq:large_iso_aniso_HL}
\begin{align}
     U_{\rm iso,large}(a) & \approx \frac{\Lambda}{64}a^4 - \frac{3}{64}a^2, \label{eq:large_potential_iso_HL}\\
    U_{\rm ani,large}(a) & \approx \frac{3}{8}a^2.
\end{align} 
\end{subequations}
The above potential resembles that in General Relativity, which can be seen by comparing the two equations \eqref{eq:WDW_GR} and \eqref{eq:WDW_HL_1}, which are the same in the large universe limit with $\lambda=1$. Like in the General Relativity case, the above isotropic potential has one turning point at $a=\sqrt{3/\Lambda}$. Note that due to the approximation, we have lost information regarding the other turning points of the complete isotropic potential \eqref{eq:complete_isotropic_potential_HL}. The different choices for the coupling parameters $g_{\rm r}$ and $g_{\rm s}$ may lead to different shapes of the potential $U_{\rm iso}(a)$. As we have already stated, we are interested in the case when $U_{\rm iso}(a)$ has three distinct, real, positive roots or turning points (see \ref{fig:UV_IR_Extreme}). Further note that since $\Lambda>0$, the quantum potential $-U_{\rm iso}$ is negative to the right of the rightmost turning point. As a result, the other two turning points create a classically allowed confining region, where a cyclical universe is possible (see, \textit{e.g.} \cite{Mukohyama:2010xz,Misonoh:2011mn,Maeda:2010ke}). In the approximation above, not only the information regarding the two turning points is lost, but also the location of the rightmost turning point is overestimated. The above approximation then may be valid in the region $a\gg \sqrt{3/\Lambda}$.

Now, from \eqref{eq:leading}, using the potentials \eqref{eq:large_iso_aniso_HL}, we get
\begin{align}
    S_{0}(a) = \begin{cases}
    s_{1}\frac{3\pi^2 i}{2\kappa^2\Lambda} \sqrt{\frac{3\lambda-1}{2}}  \left(\frac{\Lambda}{3}a^2-1\right)^{\frac{3}{2}}, & a>\sqrt{\frac{3}{\Lambda}}, \\
    -\frac{3\pi^2}{2\kappa^2\Lambda} \sqrt{\frac{3\lambda-1}{2}}  \left(1-\frac{\Lambda}{3}a^2\right)^{\frac{3}{2}}, & a<\sqrt{\frac{3}{\Lambda}}.
    \end{cases} 
\end{align}
Here, for the region under the barrier, we have only chosen the solution with $s_{1}=+1$ since as stated before, the numerical solution with the full potential, which we will discuss in the section \ref{sec:numerical}, suggests that only $s_{1}=+1$ solutions are stable under the barrier. With this choice, we can solve \eqref{eq:riccati} and the result for $\Omega(a)$ is
\begin{align}\label{eq:riccati_sol_large_universe}
    \Omega(a)=\begin{cases}
        -\frac{\sqrt{3}\left(\frac{3\lambda-1}{2}\right)}{is_{1}\sqrt{\frac{\Lambda}{3}a^2-1} + \sqrt{8\left(\frac{3\lambda-1}{2}\right)+1}}a^2, & a>\frac{\Lambda}{3}, \\
        -\frac{\sqrt{3}\left(\frac{3\lambda-1}{2}\right)}{\sqrt{1-\frac{\Lambda}{3}a^2} + \sqrt{8\left(\frac{3\lambda-1}{2}\right)+1}}a^2, & a<\frac{\Lambda}{3}.
    \end{cases}
\end{align}
We have only retained the solutions for $\Omega$ such that the anisotropic wave function is stable/normalizable. Numerical solution of $\Omega(a)$ against the full potential given in \eqref{eq:full_potential_2} shows that the above solution for $\Omega(a)$ holds good for $a>\Lambda/3$. We will discuss this further in the following subsection.

Next, we determine the first order solution from \eqref{eq:first_order_wkb_hl} and the result is
\begin{align}
    S_{1} = \begin{cases}
        \log\left(\frac{a^{-\frac{1}{2}(1+p)}}{\left(\frac{\Lambda}{3}a^2 - 1\right)^{\frac{1}{4}}} \mathscr{V}_{s_{1}}(a) \right),  & a>\sqrt{\frac{3}{\Lambda}}, \\
        \log\left(\frac{a^{-\frac{1}{2}(1+p)}}{\left(1 - \frac{\Lambda}{3}a^2\right)^{\frac{1}{4}}} \mathscr{W}(a) \right), & a<\sqrt{\frac{3}{\Lambda}},
    \end{cases}
\end{align}
where the functions $\mathscr{V}_{s_{1}}(a)$, and $\mathscr{W}(a)$ are defined as
\begin{subequations}
    \begin{align}
    \mathscr{V}_{s_{1}}(a)& = \frac{\left(1+i\sqrt{\frac{a^2 \Lambda }{3}-1}\right)^{\alpha_{+}}\left(1-i\sqrt{\frac{a^2 \Lambda }{3}-1}\right)^{\alpha_{-}}
   }{i s_{1}
   \sqrt{\frac{\Lambda }{3}a^2-1}+\sqrt{3} \sqrt{4 \lambda -1}},\\
\mathscr{W}_{s_{1}}(a)&=\frac{\left(1+\sqrt{1-\frac{a^2 \Lambda }{3}}\right)^{\alpha_{+}}\left(1-\sqrt{1-\frac{a^2 \Lambda }{3}}\right)^{\alpha_{-}}
   }{s_{1}
   \sqrt{1-\frac{\Lambda }{3}a^2 }+\sqrt{3} \sqrt{4 \lambda -1}},
\end{align}
\end{subequations}
and the constants $\alpha_{\pm}$ are defined as
\begin{align}
    \alpha_{\pm} \equiv \frac{1}{2} \left(1 \pm s_{1} \sqrt{12 \lambda -3}\right).
\end{align}

Then the wave function for a large anisotropic universe in the Ho\v{r}ava-Lifshitz theory can be written, up to an overall normalization constant, as
\begin{widetext}
\begin{align}\label{eq:HL_WKB_large}
    \Psi_{\rm HL}^{\rm WKB}(a\gg1,\boldsymbol{\beta})\simeq \begin{cases}
        E \e^{-i\frac{\pi}{4}}\frac{a^{-\frac{1}{2}(1+p)}}{\left(\frac{\Lambda}{3}a^2 - 1\right)^{\frac{1}{4}}} \mathscr{V}_{+}(a) \, \e^{+\frac{3\pi^2 i}{2\kappa^2\Lambda\hbar} \sqrt{\frac{3\lambda-1}{2}}  \left(\frac{\Lambda}{3}a^2-1\right)^{\frac{3}{2}}-\frac{\sqrt{\frac{3\lambda-1}{2}}\frac{6\pi^2}{\kappa^2\hbar}a^2}{i\sqrt{\frac{\Lambda}{3}a^2-1} + \sqrt{8\left(\frac{3\lambda-1}{2}\right)+1}}\boldsymbol{\beta}^2} \\
        ~~~~~~~~~~~ + E \e^{+i\frac{\pi}{4}} \frac{a^{-\frac{1}{2}(1+p)}}{\left(\frac{\Lambda}{3}a^2 - 1\right)^{\frac{1}{4}}} \mathscr{V}_{-}(a) \,  \e^{-\frac{3\pi^2 i}{2\kappa^2\Lambda\hbar} \sqrt{\frac{3\lambda-1}{2}}  \left(\frac{\Lambda}{3}a^2-1\right)^{\frac{3}{2}}-\frac{\sqrt{\frac{3\lambda-1}{2}}\frac{6\pi^2}{\kappa^2\hbar}a^2}{-i \sqrt{\frac{\Lambda}{3}a^2-1} + \sqrt{8\left(\frac{3\lambda-1}{2}\right)+1}}\boldsymbol{\beta}^2}, & a>\sqrt{\frac{3}{\Lambda}}, \\
        E\frac{a^{-\frac{1}{2}(1+p)}}{\left(1 - \frac{\Lambda}{3}a^2\right)^{\frac{1}{4}}} \mathscr{W}_{+}(a) \, \e^{-\frac{3\pi^2}{2\kappa^2\Lambda} \sqrt{\frac{3\lambda-1}{2}}  \left(1-\frac{\Lambda}{3}a^2\right)^{\frac{3}{2}}-\frac{\sqrt{\frac{3\lambda-1}{2}}\frac{6\pi^2}{\kappa^2\hbar}a^2}{\sqrt{1-\frac{\Lambda}{3}a^2} + \sqrt{8\left(\frac{3\lambda-1}{2}\right)+1}}\boldsymbol{\beta}^2}, & a<\sqrt{\frac{3}{\Lambda}}. 
    \end{cases}
\end{align}
\end{widetext}
Notice that $s_{1}=+1$ has been chosen under the potential barrier to be consistent with the numerical solution (see section \ref{sec:numerical}). Here, we have matched the coefficients by analytically continuing from the classically forbidden region to accessible region as $(1-\frac{\Lambda}{3}a^2) \to (\frac{\Lambda}{3}a^2 - 1) \e^{\pm i \pi}$. Notice that the above wave function coincides with the Hartle-Hawking no-boundary wave function in the appropriate limit
\begin{align}
    \Psi_{\rm HL}^{\rm WKB}(a\gg 1,\boldsymbol{\beta}) \Big|_{\lambda=1} & = \alpha \, \Psi_{\rm GR}^{\rm (HH)}(a,\boldsymbol{\beta}),
\end{align}
where $\alpha = E/C$ is the ratio between the normalization constants in the two theories. Therefore, we see that in the large universe or IR limit, along with $\lambda=1$, predictions regarding the universe from General Relativity and Ho\v{r}ava-Lifshitz theory are the same.

The shortcoming of $\Psi_{\rm HL}^{\rm WKB}(a\gg1,\boldsymbol{\beta})$, however, is that the approximation we have implemented loses information regarding the isotropic potential in the intermediate region, where analytical solution for the Riccati equation is difficult to come by and numerical analysis must be performed.

In the following section, we discuss numerical solution of the Riccati equation \eqref{eq:riccati} with the complete form of the potential \eqref{eq:full_potential_2}.
\subsection{Numerical solution to the Riccati equation}\label{sec:numerical}
The form of the potentials \eqref{eq:full_potential_2} is fairly complicated and an analytical solution for the $\Omega(a)$ can not be achieved. Therefore, we turn our attention to a numerical analysis of the Riccati equation \ref{eq:riccati}. We choose the values of the parameters $\Lambda$, and $g_{i}$s such that the isotropic potential \eqref{eq:complete_isotropic_potential_HL} has three distinct real, positive roots. This case is particularly interesting because there is a classically allowed confining region where oscillating/cyclic universe solutions are possible, and quantum mechanically, a tunneling phenomenon from the oscillatory region to a forever expanding solution may be studied. For our analysis, we pick $\Lambda$, and $g_{i}$s quite generically, and even though we demonstrate, here, the analysis for one set of values for the parameters, we expect that the qualitative conclusions drawn from this analysis is sufficiently generic to include a large class of parameter values for which also the potential has three real, positive roots. Further, we have set $\lambda=1$, for the analysis, and generalizing the analysis for other values $\lambda$ is also possible. The parameter values chosen for the numerical analysis is as follows
\begin{equation}
    \begin{gathered}
        \kappa=\hbar=1, ~~~ \Lambda = 0.2,\\
        g_{\rm r} = 1.1, ~~~ g_{\rm s}= -0.05,\\
        g_{A} = -0.04, ~~~ g_{B}=-0.01.
    \end{gathered}
\end{equation}

Let us assume that $a_{1},a_{2}$, and $a_{3}$ are three distinct real roots of the potential $U_{\rm iso}(a)$, such that $a_{1}<a_{2}<a_{3}$. The $a_{i}$s are also classical turning points (see \ref{fig:UV_IR_Extreme}). We then consider the following regions
\begin{align}
    &\text{Region 1:} ~~~~ && a<a_{1} \nonumber\\
    &\text{Region 2:} ~~~~ && a_{1}<a<a_{2} \nonumber\\
    &\text{Region 3:} ~~~~ && a_{2}<a<a_{3} \nonumber\\
    &\text{Region 4:} ~~~~ && a_{3}<a \nonumber
\end{align}
The Regions $1$ and $3$ are classically forbidden, while Regions $2$ and $4$ are classically allowed. We denote the solutions of the Riccati equation \eqref{eq:riccati} in different regions by $\Omega_{\rm sol}^{i \pm}$, where $i=1,2,3,4$ marks the region and $\pm$ stands for the choices $s_{1}=\pm 1$. We will start with region $1$, and then continue to solve in the consecutive regions to the right, where in each of the following regions the initial condition must be given by the solution in the preceding region to ensure continuity. We use the \texttt{NDSolve} solver in \texttt{Mathematica} for our analysis with the particular pre-defined method ``\texttt{StiffnessSwitching}'' and further, we have also reproduced the same results using the method ``\texttt{ExplicitRungeKutta}''.

\begin{figure*}
    \centering
    \begin{minipage}{0.5\textwidth}
        \centering
        \includegraphics[width=0.95\linewidth]{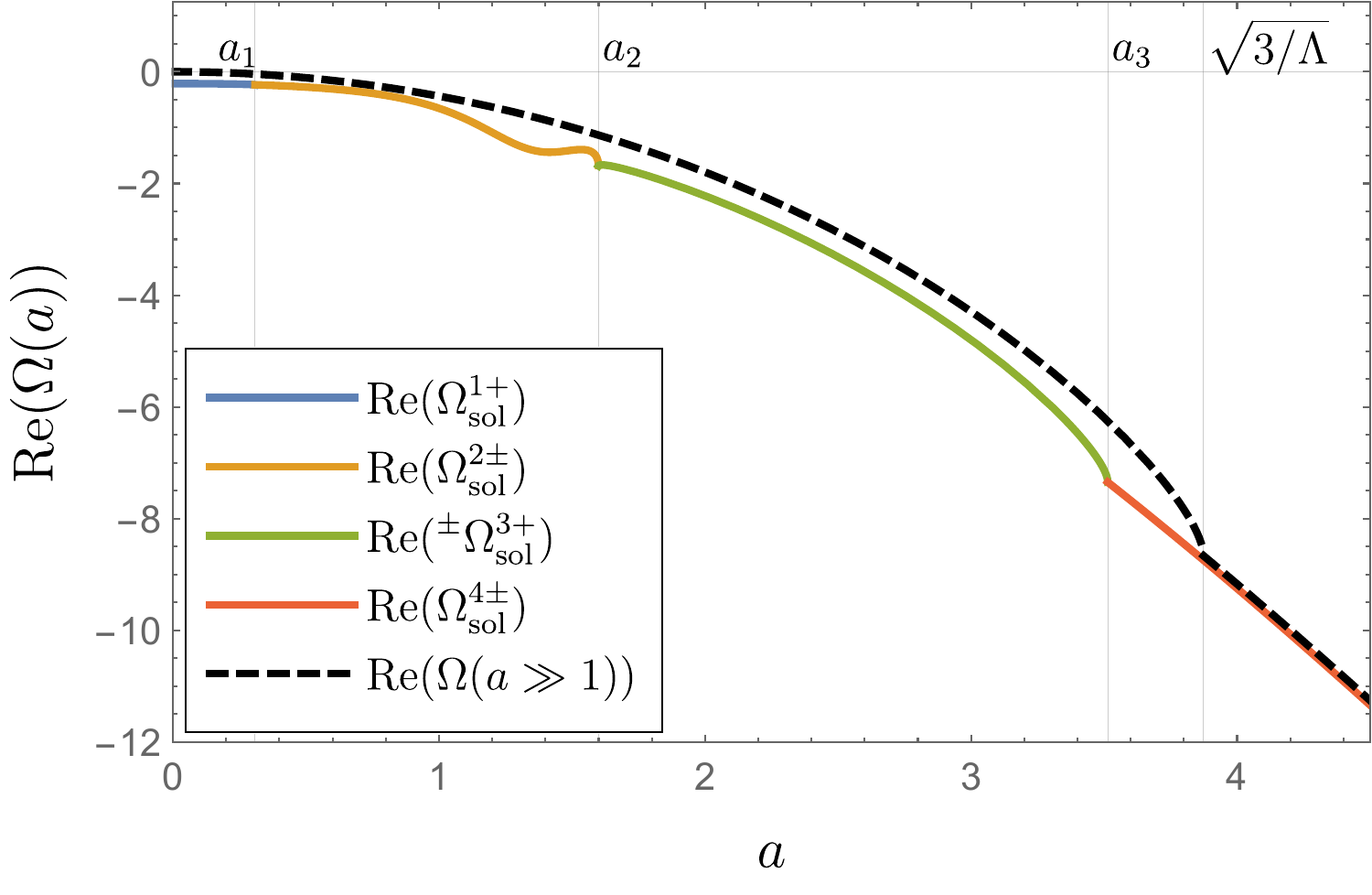}
    \end{minipage}%
    \begin{minipage}{0.5\textwidth}
        \centering
        \includegraphics[width=0.95\linewidth]{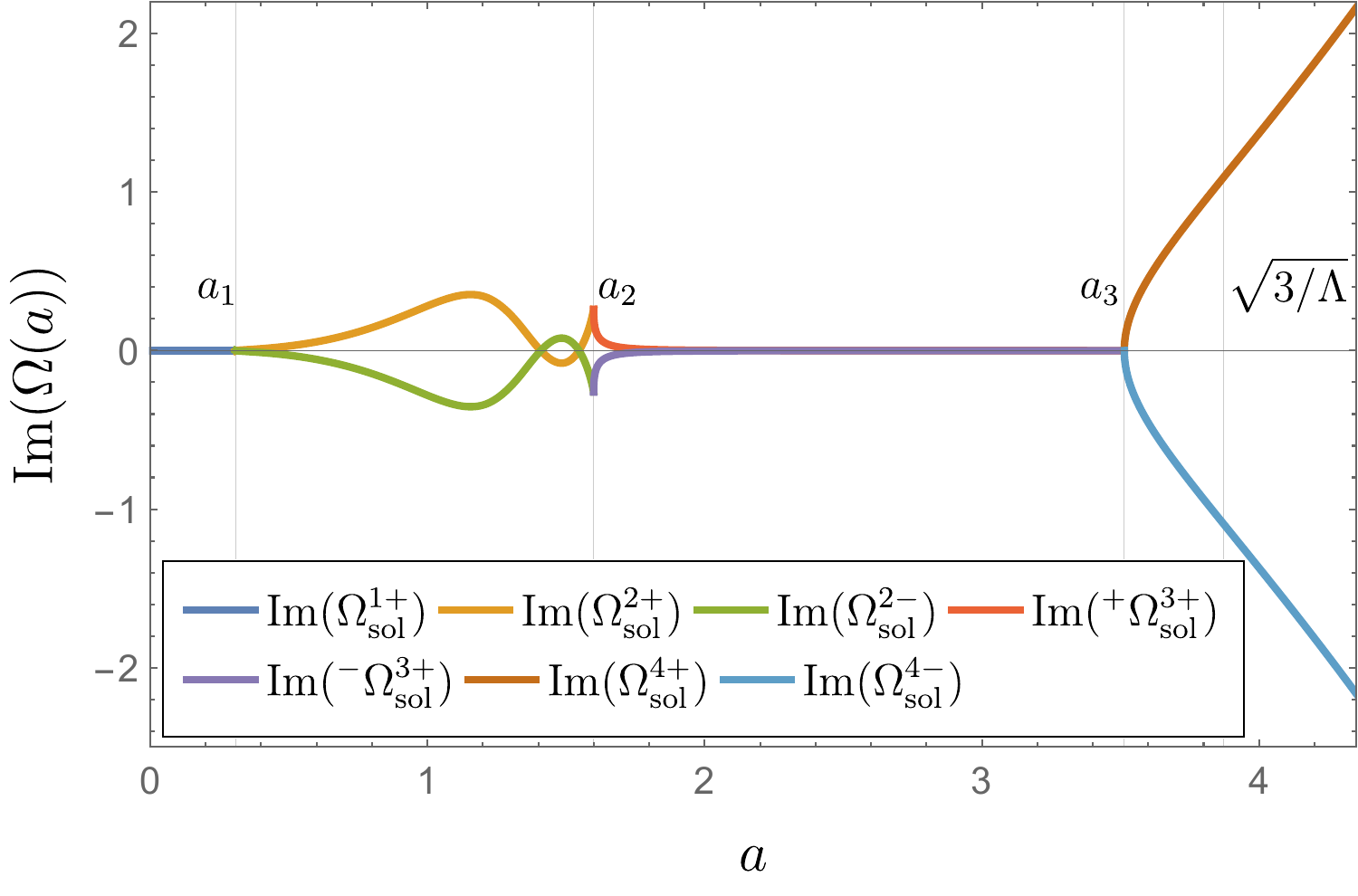}
    \end{minipage}
\caption{We consider the numerical solutions for the Riccati equation \eqref{eq:riccati}. The meaning of notations ${}^{\pm}\Omega_{\rm sol}^{i\pm}(a)$ are explained within the text. We show the real part ${\rm Re}({}^{\pm}\Omega_{\rm sol}^{i\pm}(a))$ (on the left) and ${\rm Im}({}^{\pm}\Omega_{\rm sol}^{i\pm}(a))$ (on the right) of the solutions, which satisfy the approximate matching conditions close to the turning points. We find that if in Region $1$ and $3$ (the classically forbidden regions) the $s_{1}=+1$ modes are chosen, then ${\rm Re}(\Omega(a))<0$ in all the regions indicating that the perturbation is stable and the wave function for anisotropies are normalizable. By ${\rm Re}(\Omega(a\gg 1))$ we indicate the analytical solution given in \eqref{eq:riccati_sol_large_universe}. We see that the analytical solution (black dashed line) agrees well with the numerical one after $a=\sqrt{3/\Lambda}$.}\label{fig:numerical_solution}
\end{figure*}

In Region 1, at an initial scale factor close to zero, $a=\epsilon$, where $\epsilon$ is chosen to be a positive small number, we set the initial condition as follows
\begin{align}
    \Omega_{\rm sol}^{1\pm}(\epsilon) = - \frac{3\kappa^2}{\sqrt{2}}\sqrt{\frac{3\lambda-1}{2}} \sqrt{-g_{B}}.
\end{align}
This condition is motivated by the WKB solution for $\Omega(a)$ in the small universe ($a\ll1$) limit as in equation \eqref{eq:constant_omega_hl}. With this initial condition, we numerically solve for $\Omega_{\rm sol}^{1\pm}(a)$ in the region $1$. We have set $\epsilon$ as small as $10^{-10}$ and checked that the solution remains well-behaved.

\begin{figure*}
    \centering
    \begin{minipage}{0.5\textwidth}
        \centering
        \includegraphics[width=0.95\linewidth]{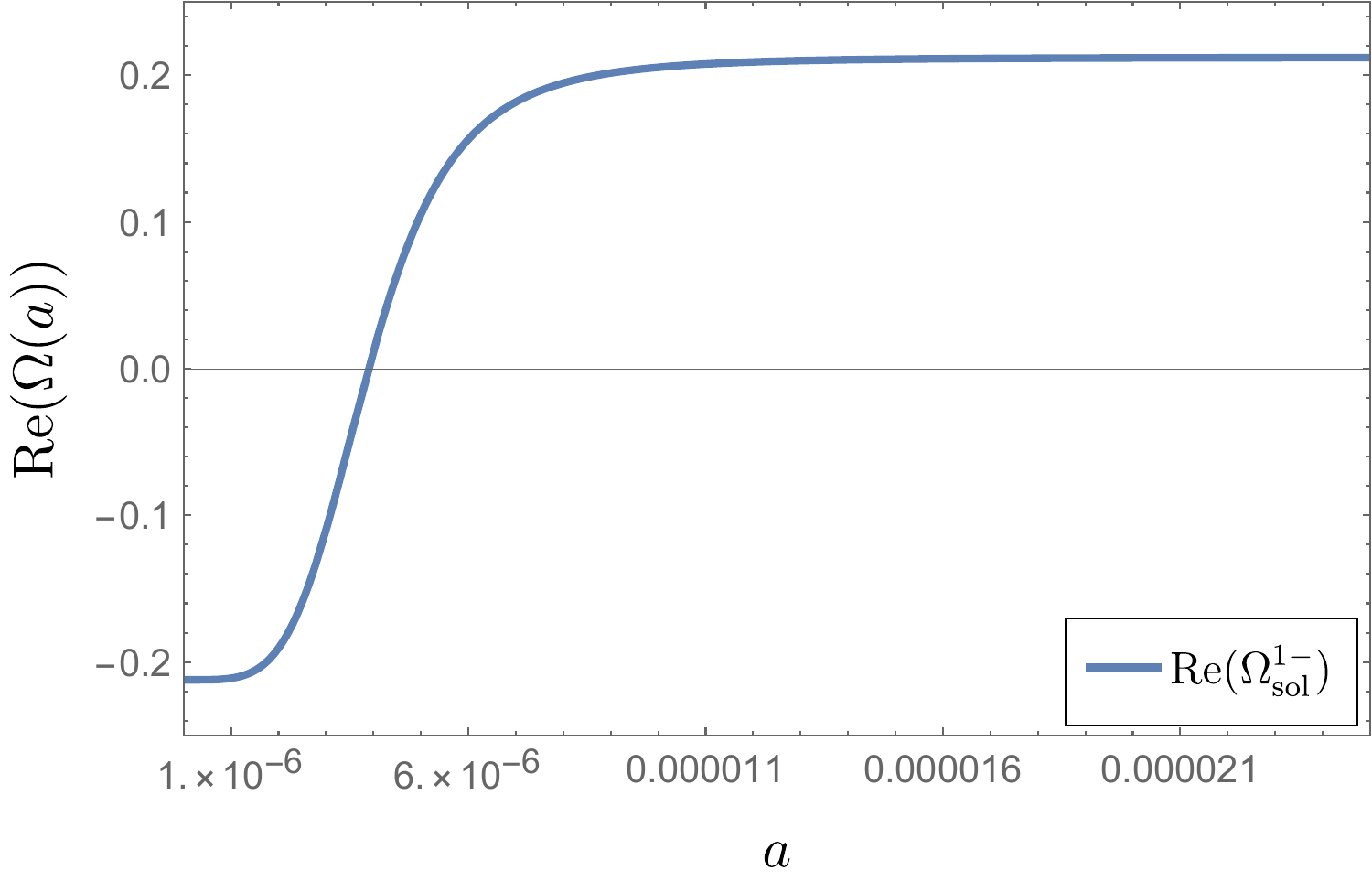}
    \end{minipage}%
    \begin{minipage}{0.5\textwidth}
        \centering
        \includegraphics[width=0.95\linewidth]{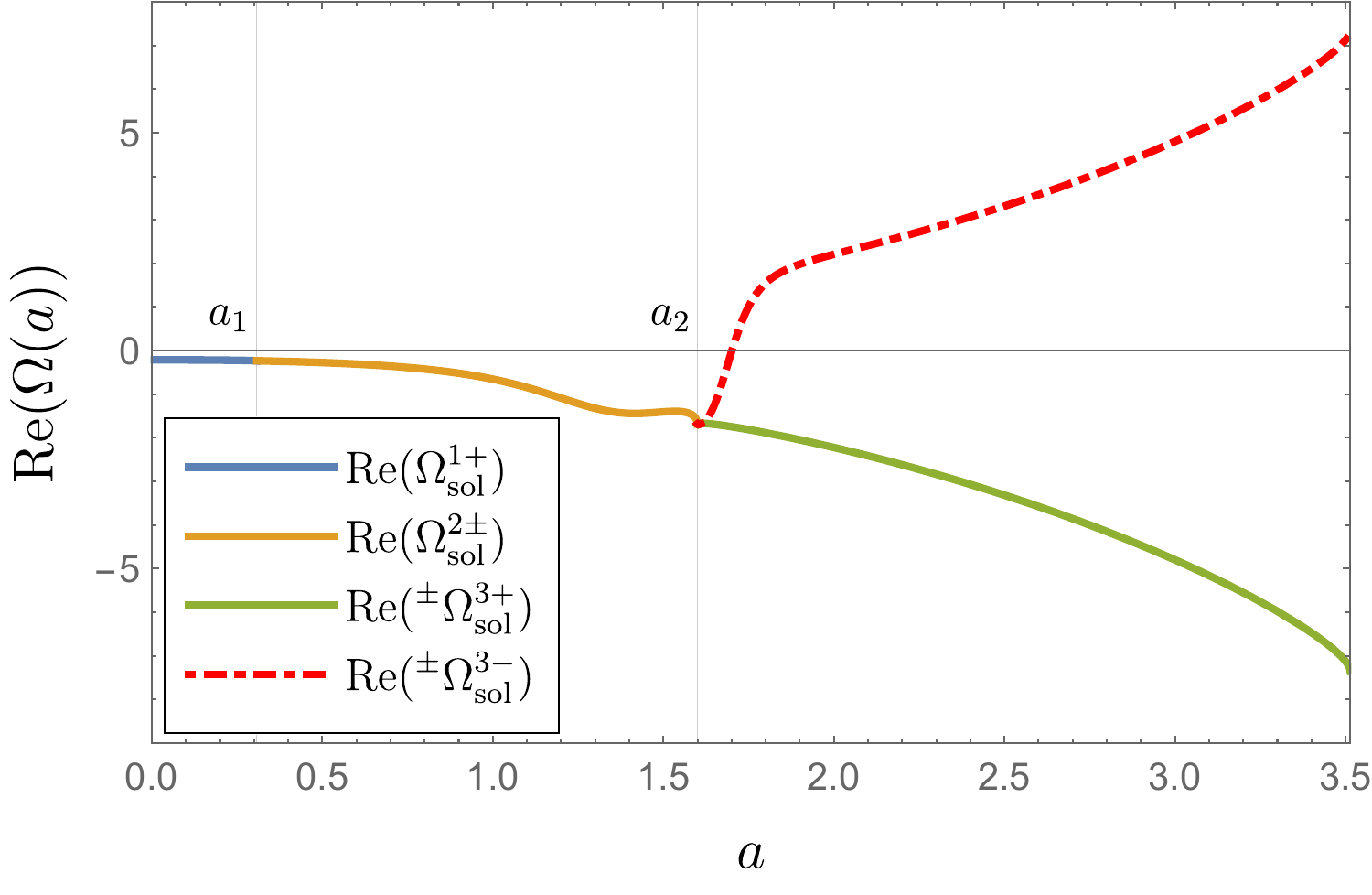}
    \end{minipage}
\caption{We plot the real part of the solutions to Riccati equations in the classically forbidden regions. In Region $1$ the solution is $\Omega_{\rm sol}^{1-}$ (left), and in Region $3$ the solution is ${}^{\pm}\Omega_{\rm sol}^{3-}$ (right), both corresponding to the choice $s_{1}=-1$. We find that the real parts flip sign leading to instability of the perturbation or non-normalizibility of the anisotropic wave functions. Thus these solutions have to discarded on physical grounds.}\label{fig:disallowed_solutions}
\end{figure*}

Numerical result shows that ${\rm Re}(\Omega_{\rm sol}^{1-})$ flips sign and becomes positive (see \ref{fig:disallowed_solutions}). For stability of the wave function, this solution must be discarded. Hence the acceptable solution in this region is $\Omega_{\rm sol}^{1+}$.

In region $2$, the initial condition for $\Omega_{\rm sol}^{2\pm}$ must be provided by the solution $\Omega_{\rm sol}^{1+}$. However, the Riccati equation \eqref{eq:riccati} has singularities at the turning points (i.e. the roots of $U_{\rm iso}(a)$) since $U(a_{i})=0$ as well as at $a=0$. This does not automatically imply that the solutions are singular at those points. In fact the solution near the turning point, say $a_{1}$, has the following asymptotic behavior
\begin{align}\label{eq:asymptotic}
\Omega(a) \simeq \begin{cases}
    \Omega_{0} + \Omega_{\frac{1}{2}}\sqrt{a_{1}-a}, & a<a_{1},\\
    \Omega_{0} \pm i \Omega_{\frac{1}{2}}\sqrt{a-a_{1}}, & a>a_{1},
\end{cases}
\end{align}
see Appendix \ref{app:sol_near_turning_point} for more details. Thus, we can see that $\Omega(a)$ is continuous, however, non-differentiable at the turning point(s). For obtaining well-behaved numerical solutions, we have to set the matching conditions at a small distance away from the turning points, just as we set the initial condition for $\Omega_{\rm sol}^{1\pm}$ at a small distance ($0<$) $\epsilon$ ($\ll 1$) away from $a=0$.
 
For region $2$, we then choose the following initial condition
\begin{align}\label{eq:matching12}
    \Omega_{\rm sol}^{2\pm}(a_{1}+\epsilon) = & \frac{\sqrt{2}\Omega_{\rm sol}^{1+}(a_{1}-\epsilon)-\Omega_{\rm sol}^{1+}(a_{1}-2\epsilon)}{\sqrt{2}-1} \nonumber\\
    & \pm i \frac{\Omega_{\rm sol}^{1+}(a_{1}-2\epsilon) - \Omega_{\rm sol}^{1+}(a_{1}-\epsilon)}{\sqrt{2}-1},
\end{align}
which is to ensure the asymptotic behavior \eqref{eq:asymptotic} in the $\epsilon\to 0$ limit. With this initial condition we solve for $\Omega_{\rm sol}^{2\pm}(a)$ in Region $2$ and observe that
\begin{align}
    {\rm Re}(\Omega_{\rm sol}^{2+}(a)) = {\rm Re}(\Omega_{\rm sol}^{2-}(a)) < 0, ~~~~~ a_{1}<a<a_{2}.
\end{align}
Further we observe that ${\rm Im}(\Omega_{\rm sol}^{2+}(a))$ and ${\rm Im}(\Omega_{\rm sol}^{2-}(a))$ are not equal except at the points where these may intersect. More specifically, we observe for the particular parameters that
\begin{align}
    {\rm Im}(\Omega_{\rm sol}^{2+}(a_{2}-\epsilon)) \neq {\rm Im}(\Omega_{\rm sol}^{2-}(a_{2}-\epsilon)).
\end{align}
Thus, while solving for $\Omega_{\rm sol}^{3\pm}(a)$, there are two possible choices of initial condition given by either a combination of $\Omega_{\rm sol}^{2+}(a_{2}-\epsilon)$ and $\Omega_{\rm sol}^{2+}(a_{2}-2\epsilon)$ or $\Omega_{\rm sol}^{2-}(a_{2}-\epsilon)$ and $\Omega_{\rm sol}^{2-}(a_{2}-2\epsilon)$. We have to consider both the choices as, in general, in Region $2$, the wave function will be given by a linear combination of both of the solutions $\Omega_{\rm sol}^{2\pm}(a)$. We denote with ${}^{+}\Omega_{\rm sol}^{3\pm}(a)$, the solution in Region $3$ that follow the solution $\Omega_{\rm sol}^{2+}(a)$ in the Region $2$, and with ${}^{-}\Omega_{\rm sol}^{3\pm}(a)$ the solution in Region $3$ that follow the solution $\Omega_{\rm sol}^{2-}(a)$ in the Region $2$. Again, we impose the following approximate matching condition consistent with the asymptotic behavior
\begin{align}\label{eq:matching23}
    {}^{\pm}\Omega_{\rm sol}^{3+}(a_{2}+\epsilon) = & \frac{\sqrt{2}\Omega_{\rm sol}^{2\pm}(a_{2}-\epsilon)-\Omega_{\rm sol}^{2\pm}(a_{2}-2\epsilon)}{\sqrt{2}-1} \nonumber\\
    & + i \frac{\Omega_{\rm sol}^{2\pm}(a_{2}-2\epsilon) - \Omega_{\rm sol}^{2+}(a_{2}-\epsilon)}{\sqrt{2}-1},\\
    {}^{\pm}\Omega_{\rm sol}^{3-}(a_{2}+\epsilon) = & \frac{\sqrt{2}\Omega_{\rm sol}^{2\pm}(a_{2}-\epsilon)-\Omega_{\rm sol}^{2\pm}(a_{2}-2\epsilon)}{\sqrt{2}-1} \nonumber\\
    & - i \frac{\Omega_{\rm sol}^{2\pm}(a_{2}-2\epsilon) - \Omega_{\rm sol}^{2+}(a_{2}-\epsilon)}{\sqrt{2}-1}.
\end{align}
We further observe that in Region $3$, the real part of the solutions ${\rm Re}({}^{\pm}\Omega_{\rm sol}^{3-}(a))$ flip the sign, and for stability, these solutions must be discarded. Whereas ${\rm Re}({}^{\pm}\Omega_{\rm sol}^{3+}(a))<0$ throughout Region $3$ and is an acceptable solution. We note that both the allowed solutions in the Region $3$ have the same value at the rightmost turning point
\begin{align}
  \frac{{}^{+}\Omega^{3+}_{\rm sol}(a_{3}-\epsilon)-{}^{-}\Omega^{3+}_{\rm sol}(a_{3}-\epsilon)}{\sqrt{\epsilon}} \to 0 \ (\epsilon\to +0),
\end{align}
providing the same initial condition for the solutions $\Omega^{4\pm}_{\rm sol}(a)$ in the $\epsilon\to 0$ limit. Again, for explicit numerical computation, we use, here, a matching condition similar to the conditions \eqref{eq:matching12}, and we avoid repeating the expression. 

Furthermore, we find that ${\rm Re}(\Omega_{\rm sol}^{4\pm})<0$. We have now found the solutions to the Riccati equation in all four regions, such that these match approximately across the turning points. However, note that the WKB approximation in the case of an isotropic universe alone is invalid at the turning points, so the approximate matching is sufficient for our analysis. Also note that the analytical solution of the Riccati equation in the large universe limit given in \eqref{eq:riccati_sol_large_universe} agrees with the numerical solution for $a>\sqrt{3/\Lambda}$. Now, we summarize the numerical findings (see \ref{fig:numerical_solution}) as follows
\begin{equation}
    \begin{gathered}
        {\rm Re}(\Omega(a))<0, ~~~
        {\rm Im}(\Omega_{\rm sol}^{2\pm})=-{\rm Im}(\Omega_{\rm sol}^{2\mp}),\\
        {\rm Im}({}^{+}\Omega_{\rm sol}^{3+})=-{\rm Im}({}^{-}\Omega_{\rm sol}^{3+}), ~~~ {\rm Im}(\Omega_{\rm sol}^{4\pm})=-{\rm Im}(\Omega_{\rm sol}^{4\mp}),
    \end{gathered}
\end{equation}
which can be shown analytically as well. 

Further, we can include $\mathcal{O}(\hbar)$ corrections by solving for the function $S_{1}(a)$ from equation \eqref{eq:first_order_wkb_hl}
\begin{align}
    S_{1}(a) =&  - \frac{1}{4} \log |U_{\rm iso}(a)| - \frac{p}{2} \log(a) \nonumber\\
    & + s_{1} \int^a  \frac{\Omega(\tilde{a}) \, \rd \tilde{a}}{a^2 \sqrt{-U_{\rm iso}(\tilde{a})} }.
\end{align}
Note that the first two terms are the same as in the isotropic case, whereas anisotropies are responsible for the last term, which we can determine by integrating the interpolating functions ${}^{\pm}\Omega_{\rm sol}^{i\pm}(a)$ generated by numerical computations piecewise across the four regions. We define the following quantities from piecewise integrations 
\begin{subequations}
\begin{align}
    \Sigma_{1}^{+}(a) & = + \int_{\epsilon^*}^{a< a_{1}}  \frac{\Omega_{\rm sol}^{1+}(\tilde{a}) \, \rd \tilde{a}}{a^2 \sqrt{-U_{\rm iso}(\tilde{a})}}, \\
    \Sigma_{2}^{\pm}(a) = & \Sigma^{+}_{1}(a_{1}-\epsilon)\pm \int_{a_{1}+\epsilon}^{a < a_{2}}  \frac{\Omega_{\rm sol}^{2\pm}(\tilde{a}) \, \rd \tilde{a}}{a^2 \sqrt{-U_{\rm iso}(\tilde{a})}}, \\
    {}^{\pm}\Sigma_{3}^{+}(a) = &  \int_{a_{2}+\epsilon}^{a < a_{3}}  \frac{{}^{\pm}\Omega_{\rm sol}^{3+}(\tilde{a}) \, \rd \tilde{a}}{a^2 \sqrt{-U_{\rm iso}(\tilde{a})} } + \Sigma^{\pm}_{2}(a_{2}-\epsilon), \\
    {}^{\pm}\Sigma_{4}^{+}(a) = & {}^{\pm}\Sigma_{3}^{+}(a_{3}-\epsilon) + \int_{a_{3}+\epsilon}^{a}  \frac{\Omega_{\rm sol}^{4+}(\tilde{a}) \, \rd \tilde{a}}{a^2 \sqrt{-U_{\rm iso}(\tilde{a})} }, \\
    {}^{\pm}\Sigma_{4}^{-}(a) = & {}^{\pm}\Sigma_{3}^{+}(a_{3}-\epsilon) - \int_{a_{3}+\epsilon}^{a}  \frac{\Omega_{\rm sol}^{4-}(\tilde{a}) \, \rd \tilde{a}}{a^2 \sqrt{-U_{\rm iso}(\tilde{a})} }.
\end{align}
\end{subequations}
Here, the $\pm$ symbols in the quantities ${}^{\pm}\Sigma_{i}^{\pm}(a)$ refer to which solutions ${}^{\pm}\Omega_{\rm sol}^{i\pm}$ are integrated over in their respective regions. Also, $\epsilon^*$ is another small number as $\Sigma_{1}^{+}(a)$ cannot be integrated from $a=0$, and in the limit $\epsilon^*\to0$, the integration fails to converge and is logarithmically divergent. Thus, the WKB approximation is not valid near $a\sim0$ as well as at turning points. (See section \ref{sec:small_exact_HL} for the exact wave function near $a=0$.) 

We can now write down the semi-classical wave function, up to order $\mathcal{O}(\hbar)$, of the anisotropic Bianchi IX universe in the Ho\v{r}ava-Lifshitz theory as 
\begin{widetext}
\begin{align}
    \Psi_{\rm HL}^{\rm WKB}(a) \simeq \begin{cases}
        \frac{A_{+}}{a^{\frac{p}{2}}(-U_{\rm iso}(a))^{\frac{1}{4}}} \e^{-\mathscr{A}\int_{a}^{a_{1}} \sqrt{-U_{\rm iso}(\tilde{a})}\rd \tilde{a} + \Sigma^{+}_{1} + \mathscr{B} \Omega_{\rm sol}^{1+}(a) \boldsymbol{\beta}^2}, & a< a_{1}, \\
        \frac{B_{+}}{a^{\frac{p}{2}}(U_{\rm iso}(a))^{\frac{1}{4}}} \e^{i\mathscr{A}\int_{a_{1}}^{a} \sqrt{U_{\rm iso}(\tilde{a})}\rd \tilde{a}+ \Sigma^{+}_{2}+\mathscr{B} \Omega_{\rm sol}^{2+}(a) \boldsymbol{\beta}^2} + \frac{B_{-}}{a^{\frac{p}{2}}(U_{\rm iso}(a))^{\frac{1}{4}}} \e^{-i\mathscr{A}\int_{a_{1}}^{a} \sqrt{U_{\rm iso}(\tilde{a})}\rd \tilde{a}+ \Sigma^{-}_{2}+\mathscr{B}\Omega_{\rm sol}^{2-}(a) \boldsymbol{\beta}^2}, & a_{1}< a < a_{2},\\
        \frac{1}{a^{\frac{p}{2}}(-U_{\rm iso}(a))^{\frac{1}{4}}}\e^{\mathscr{A}\int_{a_{2}}^{a} \sqrt{-U_{\rm iso}(\tilde{a})}\rd \tilde{a}} \left(C_{+} \e^{{}^{+}\Sigma_{3}^{+} +\mathscr{B} {}^{+}\Omega_{\rm sol}^{3+}(a) \boldsymbol{\beta}^2} + D_{+} \e^{{}^{-}\Sigma_{3}^{+}+\mathscr{B} {}^{-}\Omega_{\rm sol}^{3+}(a) \boldsymbol{\beta}^2} \right), & a_{2}< a < a_{3} \\
        \frac{1}{a^{\frac{p}{2}}(U_{\rm iso}(a))^{\frac{1}{4}}} \e^{i\mathscr{A}\int_{a_{3}}^{a} \sqrt{U_{\rm iso}(\tilde{a})}\rd \tilde{a}+\mathscr{B} \Omega_{\rm sol}^{4+}(a) \boldsymbol{\beta}^2} \left(E_{+} \e^{{}^{+}\Sigma_{4}^{+}} + F_{+} \e^{{}^{-}\Sigma_{4}^{+}} \right)\\
        ~~~~+ \frac{1}{a^{\frac{p}{2}}(U_{\rm iso}(a))^{\frac{1}{4}}} \e^{-i\mathscr{A}\int_{a_{3}}^{a} \sqrt{U_{\rm iso}(\tilde{a})}\rd \tilde{a}+\mathscr{B} \Omega_{\rm sol}^{4-}(a) \boldsymbol{\beta}^2}\left(E_{-} \e^{{}^{+}\Sigma_{4}^{-}} + F_{-} \e^{{}^{-}\Sigma_{4}^{-}} \right), & a_{3}< a
    \end{cases}
\end{align}    
\end{widetext}
where we have defined
\begin{align}
    \mathscr{A} = \sqrt{\frac{3\lambda-1}{2}}\frac{4\sqrt{3}\pi^2}{\kappa^2\hbar}, ~~~ \mathscr{B} =  \sqrt{\frac{2}{3\lambda-1}}\frac{2\sqrt{3}\pi^2}{\kappa^2\hbar},
\end{align}
and the constants $A_{\pm},B_{\pm},C_{+},D_{+},E_{\pm}$, and $F_{\pm}$ have to be matched by analytical continuation. Let us start in the Region $3$, where for the $S_{0}$ solution only the branch with $s_{1}=+1$ has been selected to avoid the instability associated with $s_{1}=-1$. Analytically continuing from the classically forbidden Region $3$ to the left in the classically allowed Region $2$ by the paths $-U_{\rm iso}(a) \to U_{\rm iso}(a) \e^{\pm i \pi}$ which avoid the turning point, we find that the solutions must be identified as follows
\begin{align}
    C_{+}\e^{-i\frac{\pi}{4}}  & \e^{{}^{+}\Sigma_{3}^{+} +\mathscr{B} {}^{+}\Omega_{\rm sol}^{3+}(a) \boldsymbol{\beta}^2}+ D_{+}\e^{-i\frac{\pi}{4}} \e^{{}^{-}\Sigma_{3}^{+}+\mathscr{B} {}^{-}\Omega_{\rm sol}^{3+}(a) \boldsymbol{\beta}^2} \nonumber\\
    & \to B_{+} \e^{i\theta_{12}+ \Sigma^{+}_{2}+\mathscr{B} \Omega_{\rm sol}^{2+}(a) \boldsymbol{\beta}^2}, \\
    C_{+} \e^{i\frac{\pi}{4}}  &\e^{{}^{+}\Sigma_{3}^{+} +\mathscr{B} {}^{+}\Omega_{\rm sol}^{3+}(a) \boldsymbol{\beta}^2} + D_{+} \e^{i\frac{\pi}{4}} \e^{{}^{-}\Sigma_{3}^{+}+\mathscr{B} {}^{-}\Omega_{\rm sol}^{3+}(a) \boldsymbol{\beta}^2} \nonumber\\
    & \to B_{-} \e^{-i\theta_{12} + \Sigma^{-}_{2}+\mathscr{B}\Omega_{\rm sol}^{2-}(a) \boldsymbol{\beta}^2}, 
\end{align}
Since we have the approximate matching conditions \eqref{eq:matching23}, and also $\Sigma_{2}^{\pm}(a_{2}-\epsilon) = {}^{\pm}\Sigma_{3}^{+}(a_{2}+\epsilon) + \mathcal{O}(\sqrt{\epsilon})$, for the wave functions to match, we must have
\begin{align}
    C_{+} \e^{-i\frac{\pi}{4}} = B_{+} \e^{i\theta_{12}}, ~~~~ D_{+} \e^{+i\frac{\pi}{4}} = B _{-} \e^{-i\theta_{12}},
\end{align}
where we have defined
\begin{align}
    \theta_{12} & = \mathscr{A}\int_{a_{1}}^{a_{2}} \sqrt{|U_{\rm iso}(\tilde{a})|}\rd \tilde{a}.
\end{align}

Similarly, we can analytically continue the solution in the classically forbidden Region $1$ to the classically allowed Region $2$ and find the following matching condition
\begin{align}
    A_{+}\e^{-i\frac{\pi}{4}} = B_{+}, ~~~~ A_{+}\e^{+i\frac{\pi}{4}} = B_{-}.
\end{align}
Now we can match the solution from Region $3$ to $4$, leading to the conditions
\begin{equation}
    \begin{gathered}
    C_{+}\e^{\theta_{23}}\e^{-i\frac{\pi}{4}} = E_{+}, ~~~~ D_{+}\e^{\theta_{23}}\e^{-i\frac{\pi}{4}} = F_{+}, \\
C_{+}\e^{\theta_{23}}\e^{+i\frac{\pi}{4}} = E_{-}, ~~~~ D_{+}\e^{\theta_{23}}\e^{+i\frac{\pi}{4}} = F_{-},
    \end{gathered}
\end{equation}
where we have defined
\begin{align}
    \theta_{23} & = \mathscr{A}\int_{a_{2}}^{a_{3}} \sqrt{|U_{\rm iso}(\tilde{a})|}\rd \tilde{a}.
\end{align}
The quantities $\theta_{12}$ and $\theta_{23}$ can be expressed in terms of the Eliptic integral functions as can be found in \cite{Bertolami:2011ka}.

Thus, finally, the wave function takes the form
\begin{widetext}
\begin{align}
    \Psi_{\rm HL}^{\rm WKB}(a) \simeq \begin{cases}
        \frac{A_{+}}{a^{\frac{p}{2}}(-U_{\rm iso}(a))^{\frac{1}{4}}} \e^{-\mathscr{A}\int_{a}^{a_{1}} \sqrt{-U_{\rm iso}(\tilde{a})}\rd \tilde{a} + \Sigma^{+}_{1} + \mathscr{B} \Omega_{\rm sol}^{1+}(a) \boldsymbol{\beta}^2}, & a< a_{1}, \\
        \frac{A_{+}\e^{-i\frac{\pi}{4}}}{a^{\frac{p}{2}}(U_{\rm iso}(a))^{\frac{1}{4}}} \e^{i\mathscr{A}\int_{a_{1}}^{a} \sqrt{U_{\rm iso}(\tilde{a})}\rd \tilde{a}+ \Sigma^{+}_{2}+\mathscr{B} \Omega_{\rm sol}^{2+}(a) \boldsymbol{\beta}^2} + \frac{A_{+}\e^{+i\frac{\pi}{4}}}{a^{\frac{p}{2}}(U_{\rm iso}(a))^{\frac{1}{4}}} \e^{-i\mathscr{A}\int_{a_{1}}^{a} \sqrt{U_{\rm iso}(\tilde{a})}\rd \tilde{a}+ \Sigma^{-}_{2}+\mathscr{B}\Omega_{\rm sol}^{2-}(a) \boldsymbol{\beta}^2}, & a_{1}< a < a_{2},\\
        \frac{A_{+}}{a^{\frac{p}{2}}(-U_{\rm iso}(a))^{\frac{1}{4}}}\e^{\mathscr{A}\int_{a_{2}}^{a} \sqrt{-U_{\rm iso}(\tilde{a})}\rd \tilde{a}} \left(\e^{i\theta_{12}} \e^{{}^{+}\Sigma_{3}^{+} +\mathscr{B} {}^{+}\Omega_{\rm sol}^{3+}(a) \boldsymbol{\beta}^2} + \e^{-i\theta_{12}} \e^{{}^{-}\Sigma_{3}^{+}+\mathscr{B} {}^{-}\Omega_{\rm sol}^{3+}(a) \boldsymbol{\beta}^2} \right), & a_{2}< a < a_{3}, \\
        \frac{A_{+}\e^{\theta_{23}}\e^{-i\frac{\pi}{4}}}{a^{\frac{p}{2}}(U_{\rm iso}(a))^{\frac{1}{4}}} \e^{i\mathscr{A}\int_{a_{3}}^{a} \sqrt{U_{\rm iso}(\tilde{a})}\rd \tilde{a}+\mathscr{B} \Omega_{\rm sol}^{4+}(a) \boldsymbol{\beta}^2} \left(\e^{i\theta_{12}} \e^{{}^{+}\Sigma_{4}^{+}} + \e^{-i\theta_{12}} \e^{{}^{-}\Sigma_{4}^{+}} \right)\\
        ~~~~+ \frac{A_{+}\e^{\theta_{23}}\e^{+i\frac{\pi}{4}}}{a^{\frac{p}{2}}(U_{\rm iso}(a))^{\frac{1}{4}}} \e^{-i\mathscr{A}\int_{a_{3}}^{a} \sqrt{U_{\rm iso}(\tilde{a})}\rd \tilde{a}+\mathscr{B} \Omega_{\rm sol}^{4-}(a) \boldsymbol{\beta}^2}\left(\e^{i\theta_{12}} \e^{{}^{+}\Sigma_{4}^{-}} + \e^{-i\theta_{12}} \e^{{}^{-}\Sigma_{4}^{-}} \right), & a_{3}< a.
    \end{cases}
\end{align}    
\end{widetext}
Therefore, we have determined the WKB wave function up to an overall normalization.

Now, we can estimate the tunneling probability for the emergence of an expanding universe to the right of the turning point $a_{3}$ (Region $4$), from an expanding universe in the Region $2$ with scale factor $a_{o}$ as 
\begin{align}
     &\mathscr{P}_{\rm HL}(a) \propto  \left(\frac{a_{o}}{a}\right)^p \left(\frac{U_{\rm iso}(a_{o})}{U_{\rm iso}(a)}\right)^{\frac{1}{2}} \e^{+2\theta_{23}}\e^{+2\mathscr{B} {\rm Re}(\Omega_{\rm sol}^{4-}(a)) \boldsymbol{\beta}^2} \nonumber\\
    & \times \e^{-2\mathscr{B} {\rm Re}(\Omega_{\rm sol}^{2-}(a_{o}))) \boldsymbol{\beta}^2} \e^{-2{\rm Re}(\Sigma_{2}^{-}(a_{o}))} \left(\e^{+2{\rm Re}({}^{+}\Sigma_{4}^{-})+2{\rm Re}({}^{-}\Sigma_{4}^{-})}\right. \nonumber \\
    & + \left. 2\e^{{\rm Re}({}^{+}\Sigma_{4}^{-}){\rm Re}({}^{-}\Sigma_{4}^{-})} \cos\left(2\theta_{12} + {\rm Im}({}^{+}\Sigma_{4}^{-})-{\rm Im}({}^{-}\Sigma_{4}^{-})\right) \right).
\end{align}
For a specific value of the scale factor of the emerging universe, the probability can, in principle, be estimated numerically from the above expression.
%
\section{Exact solution in the small universe (\texorpdfstring{$a\ll1$}{a<<1}) limit}\label{sec:small_exact_HL}

We have seen in the previous section that the WKB approximation breaks down for a small scale factor. However, unlike the General Relativity case, in the small universe limit, the Wheeler-DeWitt equation for the Ho\v{r}ava-Lifshitz theory is separable, and the equations can be solved exactly without any further approximation. Let us assume that the small universe or UV wave function has the following form
\begin{align}
    \Psi_{\rm HL}^{\rm UV}(a,\beta_{+},\beta_{-}) = \Phi(a)\xi_{+}(\beta_{+}) \xi_{-}(\beta_{-})
\end{align}
Putting the above \textit{ansatz} into equation \eqref{eq:WDW_HL_1} where the potential has been approximated as \eqref{eq:small_iso_aniso_HL}, we find the following separated equations
\begin{subequations}
\begin{align}
    & \Bigg[\hbar^2 a^2\left(\frac{\partial^2}{\partial a^2} + \frac{p}{a}\frac{\partial}{\partial a}\right) + 9\pi^4 \left(\frac{3\lambda-1}{2}\right)g_{\rm s} \nonumber\\
    & ~~~~~~~~~~~~~~~~~~~~+ E_{+}+E_{-} \Bigg] \Phi(a) = 0, \label{eq:small_iso_HL_exact} \\
    & \left[- \hbar^2 \left(\frac{\partial^2}{\partial \beta_{\pm}^2} \right) + 216\pi^4 (-g_{B}) \beta_{\pm}^2 - \frac{E_{\pm}}{\left(\frac{3\lambda-1}{2}\right)} \right] \xi_{\pm}(\beta_{\pm}) = 0, \label{eq:small_ani_HL_exact}
\end{align}
\end{subequations}
where $E_{\pm}$ are the separation constants. It is clear from \eqref{eq:small_ani_HL_exact} that in the small anisotropy limit, the anisotropies behave as quantum harmonic oscillators, with the additional assumptions that $E_{\pm}>0$, and $g_{B}<0$. The quantity $E_{+}+E_{-}$ in \eqref{eq:small_iso_HL_exact} can be thought of as a backreaction of the anisotropies on the isotropic sector.

The anisotropic wave functions $\xi_{\pm}(\beta_{\pm})$ have the usual form as the quantum harmonic oscillator wave function
\begin{align}\label{eq:harmonic_hl}
    \xi_{\pm}(\beta_{\pm}) = \frac{1}{\sqrt{2^n n!}}\left(\frac{\omega}{\pi\hbar}\right)^{\frac{1}{4}} \e^{-\frac{\omega}{2\hbar}\beta_{\pm}^2}H_{n}\left(\sqrt{\frac{\omega}{\hbar}}\beta_{\pm}\right),
\end{align}
where
\begin{align}
    \omega = 6\sqrt{6}\pi^2\sqrt{-g_{B}},
\end{align}
and the energies are quantized as
\begin{align}
    E_{\pm}^{n} = \hbar \omega \left(\frac{3\lambda-1}{2}\right) \left(n + \frac{1}{2}\right).
\end{align}
The solution for the isotropic wave function reads as follows
\begin{align}
    \Phi(a) = a^{\frac{1}{2}-\frac{p}{2}}\left(N_{+} a^{c_{+}} + N_{-} a^{c_{-}}\right),
\end{align}
where $N_{\pm}$ are the integration constants and the powers $c_{\pm}$ are defined as
\begin{align}
    c_{\pm} = &  \pm \frac{1}{2}\Big(-18\pi^4(3\lambda-1) g_{\rm s}-2\hbar(m+n+1)(3\lambda-1)\omega \nonumber\\
    & +(p-1)^2\hbar^2\Big)^{\frac{1}{2}},
\end{align}
here $m,n$ are positive integers denoting the energy state of $\beta_{\pm}$. For the reality of the wave function, one may demand
\begin{align} \label{eq:reality}
    & -18\pi^4(3\lambda-1) g_{\rm s}-2\hbar(m+n+1)(3\lambda-1)\omega \nonumber\\
    & ~~~~~~~~~~ +(p-1)^2\hbar^2 >0,
\end{align}
which puts a bound on the coupling parameters of the Ho\v{r}ava-Lifshitz theory. Further, the DeWitt criterion $\Phi(0)=0$, which is often posed as the criterion for singularity resolution since the wave function and thus probability vanish at the singular configuration, is satisfied if
\begin{align}\label{eq:dewitt}
    \frac{1}{2}-\frac{p}{2}+c_{\pm}>0.
\end{align}
Thus the particular form of the wave function depends on the coupling parameters of the Ho\v{r}ava-Lifshitz theory and the choice of operator ordering. Previously, it has been shown that DeWitt criterion is consistent with Ho\v{r}ava-Lifshitz theory when inhomogeneous perturbations to the isotropic background are included, whereas General Relativity appears to be inconsistent with the criterion as the wave function for the perturbations is not normalizable \cite{Matsui:2021yte,Martens:2022dtd}. This conclusion is supported in the present analysis as well. We can see that the anisotropic wave function in the Ho\v{r}ava-Lifshitz theory \eqref{eq:harmonic_hl} is normalizable and well defined in the limit $a\to0$. Whereas in General Relativity the anisotropic wave functions \eqref{eq:tunneling_wave_function} and \eqref{eq:HH_anisotropic} are not well defined in the limit $a\to 0$, as the Gaussian has a diverging spread. Also note that both the Hartle-Hawking no-boundary and Vilenkin's tunneling wave functions do not satisfy the DeWitt criterion.

Now, since our solutions are exact and not perturbatively expanded in $\hbar$, the operator ordering does play an important role. For values of various parameters, if both the conditions \eqref{eq:reality} and \eqref{eq:dewitt} are satisfied, then the DeWitt criterion is not strong enough to determine the constants $N_{\pm}$. However, in some cases, at least one of these constants can be determined. For demonstration, we may choose $p=1$, which corresponds to the Laplace-Beltrami ordering. In this case, for the DeWitt criterion to be satisfied, we must have $N_{-} = 0$, and the other constant must be related to the overall normalization of the wave function and cannot be determined from an approximate solution in a particular regime.

Since the anisotropic wave functions \eqref{eq:harmonic_hl} are exact up to all orders of $\hbar$, unlike the WKB approximation and further since $\Psi^{\rm UV}_{\rm HL}(a,\beta_{\pm})$ remains well defined at $a=0$, unlike in General Relativity case, therefore $\Psi^{\rm UV}_{\rm HL}(a,\beta_{\pm})$ is well-suited to study the initial condition for anisotropies as we will discuss in the next section.
\section{Intial conditions and anisotropic shear}\label{sec:initial_conditions}
We now turn our attention to the physical predictions from the wave functions in different regimes. First, let us address the issue of the initial condition of the anisotropies in the UV or small universe limit. It is reasonable to assume that the anisotropies in the beginning started in their ground state
\begin{align}
    \xi_{\pm}(\beta_{\pm}) = \left(\frac{\omega}{\pi\hbar}\right)^{\frac{1}{4}} \e^{-\frac{\omega}{2\hbar}\beta_{\pm}^2}
\end{align}
The ground state wave function is of Gaussian form and independent of the scale factor or the physical size of the universe. Further, note that the energy gap between this ground state and the first excited state is given by
\begin{align}
    \Delta E_{\pm} = \hbar \omega \left(\frac{3\lambda-1}{2}\right).
\end{align}
Since in the UV regime it is expected that $\lambda\gg1$ \cite{Gumrukcuoglu:2011xg}, the ground state and the first excited state remain well separated, making it less likely for the anisotropies to make a transition to excited states from the ground state.

First, notice that due to the Gaussian nature of the anisotropic ground state, we find that the expectation values for the anisotropies vanish identically $\langle \beta_{\pm}\rangle = 0$. Since expectation values correspond to classical quantities due to the Ehrenfest theorem, the initial classical anisotropies should vanish. However, the quantum fluctuations around this classical value are characterized by the expectation value of the squared anisotropies.

In the UV regime, the expectation values for squared anisotropies are
\begin{align}
    \langle\hat{\beta}_{\pm}^2\rangle_{\rm UV} &  = \frac{\int_{-\infty}^{\infty} \rd^2\boldsymbol{\beta} \, \beta_{\pm}^2 (\Psi_{\rm HL}^{{\rm UV}}(a,\beta_{\pm}))^*\, \Psi_{\rm HL}^{{\rm UV}}(a,\beta_{\pm})}{\int_{-\infty}^{\infty} \rd^2\boldsymbol{\beta} \, (\Psi_{\rm HL}^{{\rm UV}}(a,\beta_{\pm}))^* \, \Psi_{\rm HL}^{{\rm UV}}(a,\beta_{\pm})} \nonumber\\
    & = \frac{\hbar}{12\sqrt{6}\pi^2\sqrt{-g_{B}}}.
\end{align}
Thus, the quantum theory predicts that the amount of initial anisotropies in the universe was probably around the characteristic scale
\begin{align}
    \beta_{\pm}^{\rm cl} \equiv \sqrt{\langle\hat{\beta}_{\pm}^2\rangle_{\rm UV}} = \frac{\sqrt{\hbar}}{2\sqrt{3}\pi (-6g_{B})^{\frac{1}{4}}}.
\end{align}
This would explain why the universe started with small anisotropies, unless the coupling parameter $|g_{B}|$ is fine-tuned to take a value close to zero. In principle, from the observations, a bound on the coupling parameters may be obtained.

We can also define the anisotropic shear as
\begin{align}
    \sigma^2 \equiv \frac{3}{N^2} (\dot{\beta}_{+}^2 + \dot{\beta}_{-}^2) =  \frac{4\kappa^4}{3\pi^4 a^6}(p_{\beta_{+}}^2 + p_{\beta_{-}}^2),
\end{align}
where we have used the definitions of the conjugate momenta \eqref{eq:conjugate_momenta_HL}. We can define an anisotropic shear operator $\hat{\sigma}^2$ by promoting the conjugate momenta to operators $p_{\beta_{\pm}} \to -i\hbar \partial_{\beta_{\pm}}$. Then the expectation value for the anisotropic shear at a hypersurface of $a={\rm constant}$ is given by
\begin{align}
    \langle\hat{\sigma}^2\rangle_{\rm UV} &  = \frac{\int_{-\infty}^{\infty} \rd^2\boldsymbol{\beta} (\Psi_{\rm HL}^{{\rm UV}}(a,\beta_{\pm}))^*\,\hat{\sigma}^2 \Psi_{\rm HL}^{{\rm UV}}(a,\beta_{\pm})}{\int_{-\infty}^{\infty} \rd^2\boldsymbol{\beta} \, (\Psi_{\rm HL}^{{\rm UV}}(a,\beta_{\pm}))^* \, \Psi_{\rm HL}^{{\rm UV}}(a,\beta_{\pm})} \nonumber\\
    & = \frac{8\sqrt{6}\kappa^4\hbar}{a^{6}\pi^2}\sqrt{-g_{B}}.
\end{align}
The anisotropic shear in the UV regime falls as $\sim a^{-6}$.

We can now determine the prediction for anisotropies and shear in the IR or the large universe limit. Note that since the numerical exploration has shown that the analytical solution for $\Omega(a)$ in \ref{eq:riccati_sol_large_universe} is in reasonable agreement with the numerical result when $a>\sqrt{3/\Lambda}$, we will use the wave function \eqref{eq:HL_WKB_large} in the IR regime. Taking only the expanding branch of the wave function, we get the following expectation values for the anisotropies (squared) at $a={\rm constant}$ hypersurfaces
\begin{align}
    \langle\hat{\beta}_{\pm}^2\rangle_{\rm IR} &  = \frac{\int_{-\infty}^{\infty} \rd^2\boldsymbol{\beta} \, \beta_{\pm}^2 (\Psi_{\rm HL}^{{\rm WKB}}(a,\beta_{\pm}))^*\, \Psi_{\rm HL}^{{\rm WKB}}(a,\beta_{\pm})}{\int_{-\infty}^{\infty} \rd^2\boldsymbol{\beta} \, (\Psi_{\rm HL}^{{\rm WKB}}(a,\beta_{\pm}))^* \, \Psi_{\rm HL}^{{\rm WKB}}(a,\beta_{\pm})} \nonumber\\
    & = \frac{\hbar\kappa^2\left(\frac{\Lambda}{3}a^2+4(3\lambda-1)\right)}{12\sqrt{6} \pi^2 \sqrt{12\lambda^2-7\lambda+1} a^2}.
\end{align}
For intermediate sizes of the universe, the expectation values vary with the scale factor, but for a very large universe, the anisotropies saturate to a constant value
\begin{align}
    \sqrt{\langle\hat{\beta}_{\pm}^2\rangle_{\rm IR}} & \stackrel{a\gg1}{\sim} \frac{\kappa\sqrt{\hbar}\sqrt{\frac{\Lambda}{3}}}{2\sqrt{3} \pi (72\lambda^2-42\lambda+6)^{\frac{1}{4}}}.
\end{align}
This is expected since the WKB wave function \eqref{eq:HL_WKB_large} bears formal resemblance to the Hartle-Hawking no-boundary wave function with perturbations. Due to homogeneity, there are only two anisotropy components $\beta_{\pm}$, whereas in the inhomogeneous case, there is a tower of infinitely many modes. However, in the large universe limit, the expectation value for each perturbation mode (squared) with respect to the Hartle-Hawking wave function also assumes a constant value (see, for example \cite{di2019noprescription}).

The expectation value for the anisotropic shear is
\begin{align}
    \langle\hat{\sigma}^2\rangle_{\rm IR} &  = \frac{\int_{-\infty}^{\infty} \rd^2\boldsymbol{\beta} (\Psi_{\rm HL}^{{\rm WKB}}(a,\beta_{\pm}))^*\, \hat{\sigma}^2 \Psi_{\rm HL}^{{\rm WKB}}(a,\beta_{\pm})}{\int_{-\infty}^{\infty} \rd^2\boldsymbol{\beta} \, (\Psi_{\rm HL}^{{\rm WKB}}(a,\beta_{\pm}))^* \, \Psi_{\rm HL}^{{\rm WKB}}(a,\beta_{\pm})} \nonumber\\
    & = \frac{16\hbar\kappa^2\sqrt{12\lambda^2-7\lambda+1}}{\sqrt{6} \pi^2(4\lambda-1) a^4}.
\end{align}
We see that the anisotropic shear falls as $\sim a^{-4}$ in the IR or large universe limit.
\section{Conclusion}
In this paper, we have determined the wave function of an anisotropic Bianchi IX universe in Ho\v{r}ava-Lifshitz theory, in the limit of small anisotropies and under the assumption that the space consists of only one connected piece. We considered a particular case, where the isotropic superpotential has three turning points, allowing the possibility of a classical oscillating universe in addition to a classical expanding universe. 

First, we have solved the Wheeler-DeWitt equation with the WKB approximation, treating anisotropies perturbatively. Due to the complicated form of the superpotential, we could only solve the equation analytically under small or large universe limits. In the analytic study, due to these assumptions, the information regarding the superpotential in the intermediate region is lost. We have then also computed a numerical solution for the anisotropies without any additional assumption on the form of the isotropic and anisotropic potentials. The WKB analysis, however, is not valid for small scale factors and near the turning points.

From the WKB analyses, we have found that the anisotropic wave function in all the regions, classically accessible/forbidden, has a Gaussian form, that is, the anisotropic part of the wave function is normalizable. In the IR limit, the Ho\v{r}ava-Lifshitz wave function with $\lambda=1$ becomes equal to the Hartle-Hawking no-boundary wave function in General Relativity (up to an overall numerical factor). Further, the WKB form of the wave function allows us to estimate the semi-classical tunneling probability of the emergence of an expanding universe on the right of the rightmost turning point from the intermediate region where, due to the confining potential, a classical oscillating universe is possible. A quantum oscillating universe confined in the intermediate region might seem impossible, as was claimed in \cite{Damour:2019iyi}. However, the system considered in \cite{Damour:2019iyi} based on a cyclic universe model in General Relativity and the one studied in the present paper based on Ho\v{r}ava-Lifshitz theory are quite different, and thus there is no contradiction. 

In Ho\v{r}ava-Lifshitz theory we have also performed analysis beyond the WKB approximation. Unlike General Relativity, in the very small universe limit, the Wheeler-DeWitt equation can be separated into isotropic and anisotropic parts. As a result, the anisotropic wave function becomes independent of the scale factor, and the isotropic wave function is consistent with the DeWitt criterion. These wave functions are accurate up to all orders of $\hbar$ and is valid at $a=0$, unlike the WKB analysis.

We have found that the quantum theory predicts that the anisotropic shear falls as $\sim a^{-6}$ in the UV regime, whereas it only falls as $\sim a^{-4}$ in the IR regime. In the IR regime, for very large scale factor, the expectation value of squared anisotropies saturates to a constant depending on the cosmological constant and the Ho\v{r}ava-Lifshitz parameter $\lambda$.

Furthermore, we see that the UV wave function predicts quantum fluctuations of anisotropies around the zero value, the size of the fluctuation being characterized by the coupling parameters of higher-dimensional operators in the Ho\v{r}ava-Lifshitz theory. Therefore, the quantum wave function of the anisotropic universe does indeed suggests that the universe started with a small amount of anisotropies.

In the present paper, we have performed a (semi-)analytical analysis with and without the WKB approximation. The WKB wave function is not valid in the limit $a\to0$. On the other hand, the study beyond the WKB approximation has allowed us to determine the wave function in the UV regime ($a\to0$). In principle, these two wave functions should be matched, which is difficult in the present case since the regimes of validity of them do not overlap with each other. Ideally, one should numerically solve the full second-order partial differential equation \eqref{eq:WDW_HL_1}, in a similar manner as in \cite{Martens:2022dtd}, to obtain the full wave function valid in all regions, instead of the (semi-)analytic approach we have taken here. With such a wave function, which connects both UV and IR regimes, it is in principle possible to put direct bounds on the quantum gravitational parameters $g_{i}$s and $\lambda$ from cosmological observations. We plan to revisit this important problem in a future project.

\section*{acknowledgements}

The research of VM is funded by the INSPIRE fellowship from the DST, Government of India (Reg. No. DST/INSPIRE/03/2019/001887). The work of SM was supported in part by JSPS (Japan Society for the Promotion of Science) KAKENHI Grant No.\ JP24K07017 and World Premier International Research Center Initiative (WPI), MEXT, Japan.

\appendix

\section{Solution of the Riccati equation near the turning points}\label{app:sol_near_turning_point}
In the following, we investigate the asymptotic behavior of the solution to the Riccati equation \eqref{eq:riccati} near the turning points $a_{i}$.

For demonstration, we take the following \textit{ansatz} for the series solution of $\Omega(a)$ near the turning point $a_{1}$
\begin{align}
\Omega(a) \simeq \begin{cases}
    \Omega_{0} + \Omega_{\frac{1}{2}}\sqrt{a_{1}-a}+\Omega_{1}(a_{1}-a) \\
    ~~~~~~ + \Omega_{\frac{3}{2}}(a_{1}-a)^{\frac{3}{2}}+\cdots, & a<a_{1},\\
    \Omega_{0} + i \Omega_{\frac{1}{2}}\sqrt{a-a_{1}}-\Omega_{1}(a-a_{1}) \\
    ~~~~~~ - i \Omega_{\frac{3}{2}}(a-a_{1})^{\frac{3}{2}}+\cdots, & a>a_{1},
\end{cases} 
\end{align}
where we have chosen the principal branch for the square root. This is because, by defining $x\equiv \sqrt{a_1-a}$ for $a<a_1$ and $x\equiv \sqrt{a-a_1}$ for $a>a_1$, the Riccati equation is rendered manifestly regular in the limit $a\to a_1-0$ and $a\to a_1+0$ in each region. Now we can expand the various the terms involved in the Riccati equation \ref{eq:riccati} near the turning point $a_{1}$. 
Then the Riccati equation \eqref{eq:riccati}, order by order, determines $\Omega_{\frac{1}{2}}$, and $\Omega_{1}$ and $\Omega_{\frac{3}{2}}$, in terms of $\Omega_{0}$, which is an integration constant. Finally, $\Omega_{0}$ should be fixed by the matching condition of the solutions in regions $a<a_{1}$ and $a_{1}<a<a_{2}$ at the turning point $a_{1}$. We do not list the solutions here since we do not need the explicit expressions for further analysis, as we are only interested in the asymptotic behavior near the turning point(s).

A similar analysis can be carried out for solutions around the other turning points $a_{2,3}$ as well.

\bibliographystyle{apsrev4-1}
\bibliography{reference}
\end{document}